\documentclass[12pt,a4paper]{article}\usepackage[]{graphicx}\usepackage[]{color}\usepackage{subfigure}
\makeatletter
\def\maxwidth{ %
  \ifdim\Gin@nat@width>\linewidth
    \linewidth
  \else
    \Gin@nat@width
  \fi
}
\newtheorem{myth}{Theorem}[section]
\newtheorem{myle}{Lemma}[section]

\definecolor{fgcolor}{rgb}{0.345, 0.345, 0.345}
\makeatletter
 {\par\unskip\endMakeFramed%
 \at@end@of@kframe}
\makeatother

\definecolor{shadecolor}{rgb}{.97, .97, .97}
\definecolor{messagecolor}{rgb}{0, 0, 0}
\definecolor{warningcolor}{rgb}{1, 0, 1}
\definecolor{errorcolor}{rgb}{1, 0, 0}
\usepackage{alltt}
\usepackage[utf8]{inputenc}
\usepackage{amsmath}
\usepackage{amsfonts}
\usepackage{amssymb}

\usepackage{graphicx}

\author{Kelly C. M. Gon\c{c}alves\\ IM-UFRJ \and Helio S. Migon\\ IM-UFRJ \and Leonardo S. Bastos\\ PROCC-Fiocruz }
\title{Dynamic quantile linear models: a Bayesian approach}
\date{}

\newcommand{\bfFt}{\mathbf{F}_t}
\newcommand{\bfGt}{\mathbf{G}_t}
\newcommand{\bfWt}{\mathbf{W}_t}

\newcommand{\bftheta}{\boldsymbol{\theta}}
\newcommand{\bfTheta}{\boldsymbol{\Theta}}

\newcommand{\mathD}{\mathcal{D}}

\begin{document}

\maketitle

\begin{abstract}

A new class of models, named dynamic quantile linear models, is presented. 
It combines dynamic linear models with distribution free quantile regression producing a robust statistical method. 
Bayesian inference for dynamic quantile linear models can be performed using an efficient Markov chain Monte Carlo algorithm. 
A fast sequential procedure suited for high-dimensional predictive modeling applications with massive data, in which the generating process is itself changing overtime, is also proposed. 
The proposed model is evaluated using synthetic and well-known time series data. The model is also applied to
predict annual incidence of tuberculosis in Rio de Janeiro state for future years and compared with global strategy targets set by the World Health Organization. 

\end{abstract}


\section{Introduction}
\label{sec:intro}

This paper aims to combine two innovative areas  developed during the last quarter of the twentieth century building a useful broad new class of models, namely dynamic linear models and quantile regression. 
In a collection of papers in the 1970s some new ideas to model time series data were put forward  by Jeff Harrison and co-authors \cite{harrison1976bayesian}. 
This class of models can be ingeniously  viewed   as  regression models with parameters varying throughout time. 
At almost the same time, Roger Koenker introduced the quantile regression models, generalizing the $L_1$ regression, a robust procedure that has since been successfully  applied to a range of statistical models \cite{Koenker1978}.
It provides richer information on the effects of the predictors than does the traditional mean regression and it is very insensitive to heteroscedasticity and outliers, accommodating the non-normal errors often encountered in practical applications.

The inferential approaches are, nevertheless, completely distinct. 
While the first contribution follows the Bayesian paradigm, the other resorts to optimization techniques to solve the stated minimization problem, and its inference is theoretically founded on large sample theory. 
In the next paragraphs, we state the main novelties in both consolidated areas.

Dynamic linear models (DLM) are part of a broad class of models with time varying parameters, useful for modeling and forecasting time series and regression data. 
They were introduced by \cite{harrison1976bayesian} and extended to generalized linear models by \cite{west1985dynamic}. 
A hierarchical version was later introduced by \cite{gamerman1993dynamic}. 
A  deeper treatment of the subject can be found in %books such as
\cite{west1997,durbin2002simple, Migon2005, Petris2009dynamic} and \cite{prado2010time}. %, and also in a revision paper \cite{Migon2005}. 
During the last 30 years, many contributions, methodological and  computational, have been introduced. 
The advent of stochastic simulation techniques stimulated applications of the state space methodology to model complex stochastic structures, like 
dynamic spatiotemporal models \cite{gamerman2003space}, 
dynamic survival models \cite{bastos2006dynamic},
dynamic latent factor models \cite{lopes2008spatial}, and
multiscale modeling \cite{ferreira2011dynamic, fonseca2017jasa}. 
A number of papers have recently appeared on the application of DLM  to  
hydrology \cite{fernandes2009modelling}, 
intraday electricity load \cite{migon2013multivariate}, 
finance \cite{zhao2016dynamic}, 
insurance and many other areas. 
The relative ease with which Markov chain Monte Carlo (MCMC) methods can be used to obtain the posterior distributions, even in complex situations, has made Bayesian inference very useful and attractive, but at the expense of losing the sequential analysis of both states and parameters.
Moreover, some fast computing alternatives exploring analytical approximations can be seen in  \cite{da2011dynamic} and \cite{souza2016extended}. 
Another advantage of our approximation procedure is that it provides the marginal likelihood sequentially as new data arrive. 
This is essential to perform sequential model monitoring and model selection \cite{West1981}.

The quantile regression models were introduced in the 1970s by \cite{Koenker1978}. 
A simple minimization problem yielding the ordinary sample quantiles in the location model was shown to generalize naturally to the linear model, generating a broad new class of models named  ``quantile regression models''.
The $\tau$-th quantile regression, $ \tau \in (0,1),$ is defined as any solution to the minimization of the expected value of a very special loss function, usually called the check function. 
Contrary to the commonly used quadratic loss function for mean regression estimation, the quantile regression links to a special class of loss functions that have robust properties.
A good comprehensive revision of quantile regression can be found in \cite{Koenker2005}. 

This is a straightforward generalization of the minimization involved in ordinary regression.
The estimator which minimizes the sum of absolute residuals is an important special case ($\tau=1/2$), often called the $L_1$ regression or the median regression.
It is surprising how long it took to recognize that irrespective of the actual distribution of the data, Bayesian inference for quantile regression proceeds by forming the likelihood function based on the asymmetric Laplace distribution \cite{yu2001bayesian}.
 
The recent literature includes a large number of papers on foundations of quantile regression, 
including advances in nonparametric 
\cite{ gannoun2003nonparametric, cai2009nonparametric, thompson2010bayesian, kim2012semiparametric}. 
In Bayesian foundations, 
\cite{alhamzawi2012prior} discuss prior elicitation, and \cite{yang2015posterior} evaluate the asymptotic validity of posterior inference for Bayesian quantile regression. 
\cite{yue2011bayesian} add random effects to account for over-dispersion caused by unobserved heterogeneity or for correlation in longitudinal data. 
Varying covariate effects based on quantile regression is explored in \cite{peng2014shrinkage}, including a new perspective on variable selection. 
A quantile regression method for hierarchical linear models is presented in \cite{yu2015bayesian}.  
Inference in the context of quantile regression process is explored  in  \cite{hagemann2017cluster},  in a setting with a large number of  small heterogeneous clusters. 
They provide consistent estimates of the asymptotic covariance function of that process.

Quantile regression is applied to temporal and  spatially referenced data as a flexible and interpretable method of simultaneously detecting changes in several features of the distribution of some variables.
A method based on estimating the conditional distribution was given by \cite{cai2002regression} and quantile autoregression was introduced by \cite{koenker2006quantile}. 
\cite{reich2012bayesian}  developed a spatial quantile regression model that  allows the covariates to affect the entire conditional distribution, rather than just the mean, and this conditional distribution is allowed to vary smoothly from site-to-site. 
Another reference in the context of spatially referenced data is \cite{lum2012spatial}.
\cite{reich2012spatiotemporal} developed a spatiotemporal model where each spatial location has its own quantile function that evolves linearly over time, and the quantile functions are smoothed spatially using Gaussian process priors.
\cite{reich2013bayesian} proposed a semi-parametric quantile regression model for censored survival data.

Applications of quantile regression have become widespread in recent years. 
A general discussion of using quantile regression for return-based analysis was given by \cite{taylor1999quantile} and \cite{bassett2002portfolio}. 
Quantile regression for rate making in the framework of heterogeneous insurance portfolios was applied by \cite{kudryavtsev2009using}. 
A spatiotemporal quantile regression model for the analysis of emergency department-related medical expenditures was developed by \cite{neelon2015spatiotemporal}. 
A Bayesian quantile ordinal model was proposed by \cite{rahman2016bayesian} and applied to a public opinion survey on tax policies. % introduces an estimation method for quantile regression in univariate ordinal models combining the latent variable inferential framework and the normal-exponential mixture representation of the asymmetric Laplace distribution. 
A Bayesian quantile regression model for insurance company cost data, introduced by \cite{sriram2016bayesian}, enables the modeling of different quantiles of the cost distribution as opposed to just the center and also helps to estimate the cost-to-output functional relationship at a firm level by borrowing information across firms. 
\cite{li2016quantile} presented a model for residual lifetime with longitudinal biomarkers.

We extend the dynamic linear models to a new class, named dynamic quantile linear models, where a linear function of the state parameters is set equal to a quantile of the response variable at time $t$, $y_t$, similar to the quantile regression of \cite{Koenker2005}. 
This method is suited for high-dimensional predictive modeling applications with massive data in which the generating process itself changes over time. 
Our proposal keeps the most relevant characteristics of DLM such as: 
(i) all relevant information sources are used:  history, factual or subjective experiences, including knowledge of forthcoming events;
(ii) everyday forecasting is generated  by a statistical model and exceptions can be taken into account as an anticipative or retrospective base;
(iii) {\it what happened} and {\it what if} analysis are easily accommodated; 
and (iv) the model can be decomposed into independent components describing particular features of the process under study.  

We introduce the inference via Markov chain Monte Carlo (MCMC) methods, and also via an alternative approach based on normal approximations and Bayes linear estimation, as used in \cite{west1985dynamic}, which besides being computationally faster than MCMC, allows sequential analysis of the data.

The remainder of the paper is organized as follows. 
Section 2 explores in more details the dynamic quantile linear model.
Section 3 presents our efficient MCMC algorithm and the sequential approach for the dynamic quantile linear modeling. 
Section 4 illustrates the proposed method with synthetic data, and also presents some results of the well-known time series of quarterly gas consumption in the UK from 1960 to 1986 and the annual flow of the Nile River at Aswan from 1871 to 1970. 
The model is also applied to time series data on tuberculosis in Rio de Janeiro state, Brazil, from 2001 to 2015. We also predict the incidence for future years and compare the results with those of the global tuberculosis strategy targets established. 
Finally, Section 5 concludes with discussion of the paper and some possible extensions.

\section{Dynamic quantile linear model}

The $\tau$-th quantile of a random variable $y_t$ at time $t$ can be represented as a linear combination of explanatory variables, 
$$\mathbb{Q}_\tau(y_t) = \bfFt' \bftheta_t^{(\tau)},$$
where $\mathbb{Q}_\tau(y_t)$ is the $\tau$-quantile of $y_t$, formally defined as $\mathbb{Q}_\tau(y_t) = \inf \{y^* : P(y_t < y^* ) \geq \tau\}$, for $0<\tau<1$. 
$\bfFt$ is a $p \times 1$ vector of explanatory variables at time $t$, and $\bftheta_t^{(\tau)}$ is a $p \times 1$ vector of coefficients depending on $\tau$ and $t$. 
From now on, the superscript $\tau$ will be omitted in order to keep the notation as simple as possible.

For a given time $t$, \cite{Koenker1978} defined the $\tau$-th quantile regression estimator of $\bftheta_t$ as any solution of the quantile minimization problem
\begin{equation*}%\label{quantmin}
\min_{\bftheta_t}\rho_\tau\left(y_t-\bfFt' \bftheta_t \right),
\end{equation*}
where $\rho_\tau(.)$ is the loss (or check) function defined by $\rho_\tau(u) = u( \tau - I(u<0)),$ with $I(\cdot)$ denoting the indicator function.
Minimizing the loss function $\rho_\tau(\cdot)$ is equivalent to maximizing the likelihood function of an asymmetric Laplace ($\cal{AL}$) distribution, as pointed out by \cite{yu2001bayesian}. 
However, instead of maximizing the likelihood, as \cite{yu2001bayesian}, we derive the posterior distribution of the $\tau$-th quantile regression coefficients at time $t$ using the $\cal{AL}$. 
Therefore, regardless the distribution of $y_t$, it is enough  to assume that:
\begin{align}
y_t \mid \mu_t, \phi, \tau \sim \mathcal{AL}\left( \mu_t, \phi^{-1/2}, \tau \right), \quad y_t \in \mathbb{R}, \qquad t=1,2,\ldots,T, 
\end{align}
where $\mu_t = {\bfFt'} \bftheta_t \in \mathbb{R}$ is a location parameter, $\phi^{-1/2}>0$ is a scale parameter, and $\tau \in (0,1)$ is a skewness 
parameter representing the quantile of interest. 
In dynamic  modeling, one goal is also to obtain the predictive distribution. 
This can be done in a robust fashion using a grid of values of $\tau \in (0,1)$ to describe the full predictive distribution of $y_t$. %  as proposed, for example by \cite{johannes2009quantile}. 
%falta citar Johannes, Polson, Yae (2009)} -- Qual a referencia? Soh achei um pdf nao publicado.
Nevertheless, in this paper, we focus on providing precise inference about the linear predictor $\mu_t$ for each $\tau$-th quantile.

In a dynamic linear model, the states at time $t$ depend on the states at time $t-1$ according to an evolution equation $\bftheta_t = \bfGt \bftheta_{t-1} + w_t$, 
where $\bfGt$ is a $(p \times p)$ matrix describing the evolution parameters and $w_t$ is a Gaussian error with variance matrix $\mathbf{W}_t$.  
%Of course it is not difficult to introduce some robust evolution for the states as we will see in one of the examples.
Therefore the proposed dynamic quantile linear model (DQLM) is defined as
\begin{eqnarray}  \nonumber
y_t | \bftheta_t, \phi, \tau & \sim & \mathcal{AL}\left( \bfFt' \bftheta_t, \phi^{-1/2}, \tau \right), \\ \label{DQLM}
\bftheta_t |\bftheta_{t-1},  \bfWt & \sim & N_p(\bfGt \bftheta_{t-1}, \bfWt), \quad t=1,2,\ldots,T.
\end{eqnarray}

In the next section, we present two methods of inference for the proposed model. In the first one, we extend the algorithm proposed by \cite{kozumi2011gibbs} for Bayesian (static) quantile regression and propose an efficient MCMC algorithm to sample from the posterior distribution of the unknown quantities of the dynamic quantile linear model (\ref{DQLM}). 
The second one is a computationally cheaper alternative that explores analytical approximations and Bayes linear optimality. An advantage of the last one is that it provides the marginal likelihood sequentially as new data arrive.

\section{Posterior inference for the DQLM}

\subsection{Efficient MCMC algorithm}

\cite{kotz2001laplace} presented a location-scale mixture representation of the $\cal{AL}$ that allows finding analytical expressions for the conditional posterior densities of the model. 
In this way, if a random variable follows an asymmetric Laplace distribution, i.e. $y_t \mid \mu_t, \phi, \tau \sim \mathcal{AL}\left( \mu_t, \phi^{-1/2}, \tau \right)$, then we can write $y_t$ using the following mixture representation:
\begin{align}\label{mix_repr}
\begin{array}{rl}
y_t\mid \mu_t,U_t,\phi,\tau &\sim N(\mu_t + a_\tau U_t,b_\tau \phi^{-1/2} U_t),\\
U_t\mid \phi &\sim Ga(1,\phi^{1/2}),
%%% y_t\mid \mu_t,u_t,\phi,\tau &\sim N(\mu_t + a_\tau u_t,b_\tau \phi^{-1/2} u_t),\\
%%% u_t\mid \phi &\sim Ga(1,\phi^{1/2}),
\end{array}
\end{align}
where $a_{\tau}=\frac{1-2\tau}{\tau(1-\tau)}$ and $b_{\tau} =\frac{2}{\tau(1-\tau)}$ are constants that depend only on $\tau$.

Therefore, the dynamic quantile linear model (\ref{DQLM}) can be rewritten as the following hierarchical model: 
\begin{align}  
\begin{array}{rl}\label{DQLM.2}
y_t | \bftheta_t, U_t, \phi, \tau & \sim  N\left( \bfFt' \bftheta_t + a_\tau U_t, b_\tau \phi^{-1/2} U_t\right), \\
%%% y_t | \bftheta_t, u_t, \phi, \tau & \sim  N\left( \bfFt' \bftheta_t + a_\tau u_t, b_\tau \phi^{-1/2} u_t\right), \\
\bftheta_t | \bftheta_{t-1}, \mathbf{W}_t & \sim  N_{p}\left( \bfGt \bftheta_{t-1} , \bfWt \right), \\ 
U_t | \phi & \sim Ga(1,\phi^{1/2}),
%%% u_t | \phi & \sim Ga(1,\phi^{1/2}),
\end{array}
\end{align}
for $t=1,\dots,T$.
To allow for some flexibility in the model (\ref{DQLM.2}), we can even assume that $\phi^{-1/2}$ changes with time, which can be done in logarithmic scale according to a random walk or using the discounted variance learning through the multiplicative gamma-beta-gamma model \cite[chapter 10, p. 357]{west1997}.

The model is completed with a multivariate normal prior for $\bftheta_0$, $\bftheta_0\sim N_p\left(\mathbf{m}_0,\mathbf{C}_0\right)$, 
an independent inverse gamma prior for $\phi^{-1/2}$, $\phi^{-1/2} \sim IGa(n_{\phi}/2, s_{\phi}/2)$, and an inverse Wishart 
prior for $\bfWt$, $\bfWt \sim IWishart( n_w, \mathbf{S}_w)$.

The posterior distribution of the parameters in the model (\ref{DQLM.2}) is given by
\begin{align}\label{posteriorDQLM.2}
\pi(\bfTheta, \mathbf{U}, \mathbf{W}, \phi \mid \mathD_{T}) \propto &\prod_{t=1}^{T}\left[ \pi(y_t | \bftheta_t, U_t, \phi) \pi(\bftheta_t|\bftheta_{t-1},\mathbf{W}_t)\pi(U_t) \pi(\bfWt)\right]\\
%%% \pi(\bfTheta, \mathbf{u}, \mathbf{W}, \phi \mid \mathD_{T}) \propto &\prod_{t=1}^{T}\left[ \pi(y_t | \bftheta_t, u_t, \phi) \pi(\bftheta_t|\bftheta_{t-1},\mathbf{W}_t)\pi(u_t) \pi(\bfWt)\right]\\
&\pi(\phi^{-1/2}) \pi(\bftheta_0),\nonumber
\end{align}
where $\mathD_{t} = \{\mathD_{t-1} \cup I_t \cup  y_t\}$ represents all information until time $t$, for $t=1,2,\ldots,T$. 
The quantity $I_t$ represents all external information at time $t$. If there is no external information at time $t$, then $I_t = \emptyset$. 
All prior information is summarized in $\mathD_0 = I_0$ containing all hyper parameters associated with the prior distributions. The unknown quantities are defined as follows: $\bfTheta = (\bftheta_0,\bftheta_1,\ldots,\bftheta_T)$,  $\mathbf{U} = (U_1,U_2,\ldots,U_T)$, $\mathbf{W} = (\mathbf{W}_1, \mathbf{W}_2,\ldots,\mathbf{W}_T)$.

We can sample from the posterior distribution (\ref{posteriorDQLM.2}) through an MCMC algorithm. 
Our starting point is the efficient Gibbs sampler for Bayesian (static) quantile regression proposed by \cite{kozumi2011gibbs}.  
The dynamic coefficients are then sampled using a forward filtering backward sampling (FFBS) algorithm \cite{carter1994gibbs,fruhwirth1994data,shephard1994partial}. 
Theorem \ref{mcmc} displays the full conditionals and the sampling algorithm.

%\teo3.1
\begin{myth}\label{mcmc}
 Let $\Phi=(\mathbf{u},\mathbf{W}, \bfTheta, \phi, t=1...T))$. A Gibbs sampling algorithm for the dynamic quantile model in (\ref{DQLM.2}) involves two main steps: 
\begin{description}
\item{1.} First,   sample  $\phi^{-1/2}$, $u_t$, and $\bfWt$ for $t=1,\dots,T$ from their full conditional distributions:
\begin{itemize}
 \item [(i)] $\phi^{-1/2} \mid \mathD_{T},\bfTheta,\mathbf{U} \sim IGa\left(n_{\phi}^*/2, s_{\phi}^*/2\right),$ where $n_{\phi}^* = n_{\phi} + 3T$ and 
 
 $s_{\phi}^* = s_{\phi} + \displaystyle\sum_{t=1}^{T}\frac{(y_t - \bfFt' \bftheta_t - a_\tau U_t)^2}{b_\tau U_t} + 2 \sum_{t=1}^{T} U_t$.
 
 \item [(ii)] $U_t\mid \mathD_{T},\bftheta_t,\phi \sim {\cal GIG}\left( \chi^*_t, \kappa^*_t,\frac{1}{2}\right),$ where $\chi_t^* = \frac{(y_t-\bfFt'\bftheta_t)^2}{b_\tau \phi^{-1/2}}$ and $\kappa_t^* = \frac{a_\tau^2}{b_\tau \phi^{-1/2}} + \frac{2}{\phi^{-1/2}}$
 and $\cal{GIG}$ is the generalized inverse Gaussian distribution.

 \item [(iii)] $ \bfWt \mid \mathD_{T},\bftheta_t \sim IWishart \left(n_w^*, \mathbf{S}_w^*\right),$ where 
$n_w^* = n_w + p + 1$ and $\mathbf{S}_w^* = \mathbf{S}_w + \left(\bftheta_t- \bfGt \bftheta_{t-1}\right)'\left(\bftheta_t- \bfGt \bftheta_{t-1}\right)$.
\end{itemize}

 \item{2.} Next, use the FFBS method to sample from $\pi(\bftheta \mid \cdot)$: 
\begin{itemize}
\item[(i)] Forward filtering: for $t=1,\dots,T$ calculate 
\begin{align*}
\mathbf{m}_t = \mathbf{a}_t + \mathbf{R}_t \bfFt q_t^{-1} (y_t - f_t) \mbox{ and } \mathbf{C}_t = \mathbf{R}_t - \mathbf{R}_t \bfFt q_t^{-1} \bfFt'  \mathbf{R}_t,
\end{align*}
with $\mathbf{a}_t = \bfGt \mathbf{m}_{t-1}$, 
% Discount factor! -> $\mathbf{R}_t = \frac{1}{\delta}\bfGt \mathbf{C}_{t-1} \bfGt'$, 
$\mathbf{R}_t = \bfGt \mathbf{C}_{t-1} \bfGt' + \bfWt$, 
$f_t = \bfFt' \mathbf{a}_t + U_t a_\tau$ 
and
$q_t = \bfFt' \mathbf{R}_{t} \bfFt + b_\tau U_t \phi^{-1/2}.$

\item[(ii)] Backward sampling: sample $\bftheta_T \sim N_{p}(\mathbf{m}_T, \mathbf{C}_T)$ and for $t=T-1,\dots,0$ sample $\bftheta_t \sim N_{p}(\mathbf{h}_t, \mathbf{H}_t)$, where
\begin{eqnarray*}
\mathbf{h}_t = \mathbf{m}_t + \mathbf{C}_t \bfGt' \mathbf{R}_{t+1}^{-1} (\bftheta_{t+1} - \mathbf{a}_{t+1}) \mbox{ and }
\mathbf{H}_t = \mathbf{C}_t - \mathbf{C}_t \bfGt' \mathbf{R}_{t+1}^{-1} \bfGt  \mathbf{C}_t.
\end{eqnarray*}
\end{itemize}

\end{description}
\end{myth}

In place of assuming an inverse Wishart prior for $\bfWt$ it is also possible to use a discount factor $\delta \in (0,1)$ subjectively assessed, controlling the loss of information. In this case the unique difference is that ${\bf R}_t$ is recalculated according to a discount 
factor $\delta$ such as $\bfWt = \frac{1-\delta}{\delta} \bfGt \mathbf{C}_{t-1} \bfGt'$. Hence, $\mathbf{R}_t$ can be rewritten as 
$\mathbf{R}_t =  \frac{1}{\delta}\bfGt \mathbf{C}_{t-1} \bfGt'$.

The full conditional distribution of $U_t$ is obtained using Lemma \ref{gig_lemma}, which shows that the generalized inverse Gaussian distribution ($\cal{GIG}$) is conjugate to the normal distribution in a  normal location-scale mixture model.

\begin{myle}\label{gig_lemma}  
 Let ${\bf y}=(y_1,\dots,y_n)$ be a normal random sample with  likelihood function   $\pi({\bf y}| U) = \prod_{i=1}^n{N(y_i|a+bU,cU)}$ and suppose that the prior distribution for $U$  is $ \cal{GIG} \left( \chi, \kappa,\lambda\right)$. Then, the posterior distribution $U \mid {\bf y}\sim \cal{GIG}(\chi^*,\kappa^*,\lambda^*),$
 where $\chi^*= \chi+c^{-1}\sum_{i=1}^n{(y_i-a)^2}$, $\kappa^*=nb^2c^{-1} + 2\kappa$ and $\lambda^*=\lambda-n/2.$  
 \end{myle}
The proof of this result is immediate and is omitted in the text. 
Moreover, note that a $Ga(\alpha,\beta)$ distribution is a particular case of a ${\cal GIG}$ with $\chi=0$, $\kappa=2\beta$ and $\lambda=\alpha$.

\subsection{Approximated dynamic quantile linear model}\label{aproximatedDQ}
\noindent
 
On the other hand, considering that data arrive sequentially, we propose an efficient and fast sequential inference procedure obtained with a closed-form solution, 
in order to update inference on unknown parameters online.

The approach also explores the mixture representation of the $\cal{AL}$ described in (\ref{mix_repr}). 
Hence, posterior computation can be conveniently carried out using conventional Bayesian updating, conditional on the gamma random variable  $U_t$. 
We also use a normal approximation to $U_t$'s distribution in the logarithm scale, introducing explicit dynamic behavior, once again, generalizing the model presented 
in (\ref{DQLM.2}).

The normal approximation to the gamma distribution, described in \cite{bernardo1981bioestadistica}, is presented below and the proof is presented in Appendix A.
%lemma3.2
\begin{myle}\label{best_lemma} 
Using the Kullback-Leibler divergence, in a large class of transformations, we have:  
\begin{itemize}
\item[(i)] The best transformation, to approximate $\theta \sim Ga(a,b)$  for a normal distribution is
$\zeta(\theta) = log(\theta)$. Then $ \zeta \simeq N[E(\zeta), V(\zeta)]$,  where
$E(\zeta) \simeq \log\left(\frac{a}{b}\right) -\frac{1}{2a}$  and  $V(\zeta) \simeq \frac{1}{a}$.

\item[(ii)] If  $\zeta \sim N[E(\zeta), V(\zeta)]$  then  $\theta= exp(\zeta) $ is such that  $E(\theta) \simeq \exp\left[E(\zeta)+V(\zeta)/2\right]$ and 
$V(\theta)  \simeq \exp\left[2 E(\zeta)+V(\zeta)\right] V(\zeta)$.
\end{itemize}
 \end{myle}

Therefore, an approximate conditional normal dynamic quantile regression is obtained as:
\begin{align}\label{model_approximated}
\begin{array}{rl}
y_t\mid \bftheta_t,u_t,\phi  &\sim N\left({\bf F}_t'\bftheta_t + a_{\tau}\phi^{-1/2} \exp(u_t), \phi^{-1}b_\tau \exp(u_t) \right),\\
\bftheta_t\mid \bftheta_{t-1},{\bf W}_t, \phi & \sim N_p \left ( {\bf G}_t\bftheta_{t-1}, \phi^{-1} {\bf W}_t\right),\\
u_t\mid u_{t-1}, W_{u,t},\phi  &\sim N\left( u_{t-1}, \phi^{-1}W_{u,t} \right).
\end{array} 
\end{align}
where $u_t=\log(U_t)$, with $U_t\sim Ga(a,b)$. Model (\ref{model_approximated}) is more flexible than model (\ref{DQLM.2})
because it allows $u_t$ to change with time according to a random walk. Furthermore, the scale parameter here is introduced in all
the model equations.

The model is completed assuming the following independent prior distributions: 
$ \bftheta_0\sim N_p({\bf m}_{0},\phi^{-1} {\bf C}_{0})$, $u_0 \sim N(m_{u,0},\phi^{-1} C_{u,0})$ 
and  $\phi^{-1} \sim Ga(n_0/2, d_0/2).$
The model in (\ref{DQLM.2}) can be viewed as a particular case of proposal (\ref{model_approximated}) assuming that 
$W_{u,t}=0, \forall t$, $m_{u,0}=-1/2$ and $C_{u,0}=1$.

The inference procedure is described below. First we present all results conditional on both $u_t$ and $\phi$ and later we integrate out those quantities.

  \medskip
\noindent
\subsubsection{Normal conditional  model}

We start the inference procedure by exploiting the advantage that, conditional on $u_t$ and $\phi$, we have normal distributions in one-step forecast and posterior distributions of $\bftheta_t$ at time $t$, so all the properties of a normal model can be used here. 
The dependence on $u_t$ appears first due to the one-step forecast distribution at time $t$ and it will appear in the posterior distribution as time passes. 
Let us define $u_{1:t} = (u_1,\dots,u_t).$ 
Theorem \ref{Teo2} presents the main steps in the inference procedure conditional on $u_t$.
  
%Teor 3.2
\begin{myth}\label{Teo2}
Assuming that the states' posterior distribution at time $t-1$ is 
$\bftheta_{t-1}\mid  \mathD_{t-1}, u_{1:(t-1)},{\bf W}_t, \phi\sim N[{\bf m}_{t-1},\phi^{-1}{\bf C}_{t-1}]$ and the conditions defining model 
(\ref{model_approximated}), it follows  that the prior distribution of $\bftheta_{t}$ and the conditional predictive distribution for any time 
$t$, given $u_t$ and $\phi$, are respectively:
 \begin{align}
  \bftheta_{t}\mid \mathD_{t-1}, u_{1:(t-1)},{\bf W}_t, \phi &\sim N_p[{\bf a}_{t},\phi^{-1}{\bf R}_{t}],\nonumber\\
  y_t\mid \mathD_{t-1}, u_{1:t},{\bf W}_t,\phi &\sim N[f_t(u_t), \phi^{-1}q_t(u_t)],\label{forec_cond_u}
  \end{align}
  with ${\bf a}_t={\bf G}_t{\bf m}_{t-1}$ and ${\bf R}_t={\bf G}_t{\bf C}_{t-1}{\bf G}_t+ {\bf W}$, 
  $f_t(u_t)={\bf F}_t'{\bf a}_{t}+a_{\tau}\phi^{-1/2}\exp(u_t)$ and $q_t(u_t)={\bf F}_t'{\bf R}_{t} {\bf F}_t + b_{\tau}\exp(u_t).$
The  conditional joint covariance between  $y_t$ and $\bftheta_t$,  given $D_{t-1}, u_{1:t}, \phi$, is easily obtained as  ${\bf R}_t {\bf F}_t$ completing the joint normal prior for $\bftheta_t$ and $y_t$. 
Therefore, the posterior density of  $ \bftheta_t$   follows as:
% \begin{align*} \left(\begin{array}{c} \bftheta_t\\ y_t\end{array}\middle|  \mathD_{t-1},u_t,\phi\right)\sim N \left[\left(\begin{array}{c}a_{t}\\f_t(u_t)\end{array}\right), \phi^{-1}\left(\begin{array}{cc}{\bf R}_{t}&{\bf R}_{t}{\bf F}_t\\{\bf F}_t'{\bf R}_{t} & q_t(u_t)\end{array}\right)\right], \end{align*} it is possible to obtain the posterior distribution of $\bftheta_t$ at time $t$ as:
\begin{align}\label{post_thet_condU}
 \bftheta_{t}\mid \mathD_{t},u_{1:t},{\bf W}_t,\phi &\sim N_p({\bf m}_{t}(u_t),\phi^{-1}{\bf C}_{t}(u_t)),
  \end{align}
 where $ {\bf m}_{t}(u_t)={\bf a}_t+{\bf R}_t{\bf F}_tq_t(u_t)^{-1}(y_t-f_t(u_t))$ and ${\bf C}_t(u_t) = {\bf R}_t - {\bf R}_t {\bf F}_t q_t(u_t)^{-1} {\bf F}_t' {\bf R}_t.$
 \end{myth}

It is worth pointing out that mean and variance of the predictive and the state posterior distributions are functions of $u_t$, as reinforced  by the notation used. 
However, since $u_t$ is unknown for all $t$, we must find those distributions marginal on them. We will do this sequentially in the one-step forecast distribution in (\ref{forec_cond_u}) for each time $t$.

\subsubsection{Marginalizing on $u_t$}

From now on, we will rewrite the time evolution equation in model (\ref{model_approximated}) as $
\bftheta^*_t\mid \bftheta^*_{t-1},{\bf W}_t^*, \phi \sim N_{p+1} \left ( {\bf G}^*_t\bftheta^*_{t-1}, \phi^{-1} {\bf W}_t^*\right),$
where $\bftheta_t^* = \left(\bftheta_t,u_t\right)'$, 
${\bf G}_t^* = \mbox{BlockDiag}\left({\bf G}_t,1\right)$ and ${\bf W}_t^*=\mbox{BlockDiag}\left({\bf W}_t,W_{u,t}\right)$ with prior distribution given by $\bftheta_0^*\mid \mathD_0,\phi \sim ({\bf m}_0^*,\phi^{-1}{\bf C}_0^*)$, where 
${\bf m}_0^* = \left({\bf m}_0,m_{u,0}\right)'$ and ${\bf C}_0^*= \left(\begin{array}{cc}{\bf C}_0&	 {\bf \Lambda}_0\\{\bf \Lambda}_0' & C_{u,0}\end{array}\right)$.

Assuming the posterior distribution at time $t-1$, $\bftheta^*_{t-1}\mid  \mathD_{t-1},{\bf W}_t^*, \phi\sim N_{p+1}({\bf m}^*_{t-1},$
$\phi^{-1}{\bf C}^*_{t-1})$.
By evolution, it follows that $\bftheta^*_{t}\mid \mathD_{t-1},{\bf W}_t^*, \phi \sim N_{p+1}({\bf a}^*_{t},\phi^{-1}{\bf R}^*_{t})$, 
with ${\bf a}^*_t={\bf G}^*_t{\bf m}^*_{t-1}$ and ${\bf R}^*_t={\bf G}^*_t{\bf C}^*_{t-1}{\bf G}^*_t+{\bf W}_t^*$. 

In particular, we have that $u_{t}\mid \mathD_{t-1},W_{u,t},\phi\sim N(a_{u,t}, \phi^{-1} R_{u,t})$ and the result (ii) leads to this distribution in the original (gamma) scale as:
 \begin{align}\label{dist_ustar_t1}
  U_t&\mid \mathD_{t-1},W_{u,t},\phi\sim Ga(\alpha_t, \beta_t),
 \end{align}
 where $\alpha_t = \phi R_{u,t}^{-1}$ and $\beta_t=\exp(-a_{u,t})\phi R_{u,t}^{-1}.$

Thus, we have that the one-step forecast distribution in (\ref{forec_cond_u}) can be seen as a normal-gamma mean-variance mixture, with the following
different features from those stated in (\ref{mix_repr}): (i)  $U_t$ in this case is gamma distributed with shape parameter different from $1$; and (ii) $q_t(u_t)$ is a linear function of $u_t$ with non null linear coefficient. 
These comments lead us to a recent class of distributions, a variant of the $\cal{AL}$, as described in Theorem \ref{NGAL}.

%teo3.3
 \begin{myth}\label{NGAL}
  The one-step ahead forecast distribution, conditional on $\phi$ and marginalized on $u_t$ arises as the convolution of independent normal and generalized asymmetric Laplace distribution ($\cal{GAL}$). 
  It can be represented as
  $$
  y_t =\zeta_t + \epsilon_t,
  $$
  where $\zeta_t\sim \mathcal{GAL}({\bf F}_t'{\bf a}_{t},a_{\tau}\phi^{-1/2}\beta_t^{-1},b_{\tau}\phi^{-1}\beta_t^{-1},\alpha_t)$ 
   and $\epsilon_t\sim N\left(0,\phi^{-1}{\bf F}_t'{\bf R}_{t} {\bf F}_t\right).$ We will refer to this as an $\cal{NGAL}$ distribution. 
 \end{myth}
   
 A brief presentation of the $\cal{NGAL}$ distribution, its moments the characteristic function, and the proof of Theorem \ref{NGAL} are presented in Appendix B. 
 Although, the $\cal{NGAL}$ distribution suffers from a lack of closed-form expressions for its probability density and cumulative distribution functions, they can be efficiently determined using numerical integration as discussed in Appendix B. 
 In particular, the one-step ahead forecast mean and variance marginal on $u_t$, can  be easily obtained through properties of conditional mean and variance as:
\begin{align*} 
E\left(y_t\mid \mathD_{t-1},\phi\right)&=E\left[E\left(y_t\mid \mathD_{t-1},U_t\right)\mid \mathD_{t-1}\right] = {\bf F}_t'{\bf a}_t + a_{\tau}\phi^{-1/2}E\left(U_t\mid \mathD_{t-1}\right)\\
& = {\bf F}_t'{\bf a}_t + a_{\tau}\phi^{-1/2}\alpha_t/\beta_t = f_t, \mbox{ and}\\
V\left(y_t\mid\mathD_{t-1},\phi\right) &= E\left[V\left(y_t\mid \mathD_{t-1},U_t\right)\mid \mathD_{t-1}\right] + V\left[E\left(y_t\mid \mathD_{t-1},U_t\right)\mid \mathD_{t-1}\right] \\
& = \phi^{-1}\left[{\bf F}_t'{\bf R}_{t} {\bf F}_t + b_{\tau}E\left(U_t\mid \mathD_{t-1}\right)\right]+a_\tau^2\phi^{-1}V\left(U_t\mid \mathD_{t-1}\right)\\
& = \phi^{-1}\left({\bf F}_t'{\bf R}_{t} {\bf F}_t + b_{\tau}\alpha_t/\beta_t+a_\tau^2\alpha_t/\beta_t^2\right)=\phi^{-1}q_t.
\end{align*}

The recurrences for posterior mean and variance may also be derived using approaches that do not invoke the normal assumption, since they possess 
strong optimality properties that are derived when the distributions are only partially specified in terms of means and variances. The Bayes linear estimation procedure, presented in \cite[Chap. 4]{west1997}, provides an alternative estimate that can be viewed as an approximation to the optimal procedure. 
Theorem \ref{Teo3} presents the main steps in the inference procedure now marginal on $u_t$. 

\begin{myth}\label{Teo3}
The joint distribution of $\bftheta^*_t$ and $y_t$ is partially described using its first and second moments, as follows:
\begin{align*}
\left(\begin{array}{c}\bftheta^*_t\\ y_t\end{array} \bigg| \mathD_{t-1},{\bf W}^*_t,\phi\right)\sim  \left[\left(\begin{array}{c}{\bf a}_{t}^*\\f_t\end{array}\right),
  \phi^{-1}\left(\begin{array}{cc}{\bf R}^*_{t}&{\bf A}_tq_t\\q_t{\bf A}_t' & q_t\end{array}\right)\right],
\end{align*}
where ${\bf A}_t= q_t^{-1}\left(\begin{array}{c}{\bf R}_t{\bf F}_t+\phi^{-1/2}a_{\tau}\exp(a_{u,t}){\bf \Lambda}_t\\
{\bf \Lambda}_t'{\bf F}_t+\phi^{-1/2}a_{\tau}\exp(a_{u,t})R_{u,t}\end{array}\right)$ and ${\bf \Lambda}_t = {\bf G}_t{\bf \Lambda}_{t-1}.$

The joint covariance between $y_t$ and $\bftheta_t^*$,  given $D_{t-1}$ and $\phi$, is obtained using the first order Taylor approximation 
$\exp(u_t)\approx\exp(a_{u,t})[1+u_t-a_{u,t}]$. 

Through the Bayes linear estimation procedure, we get:
\begin{align}\label{post_thet_condU_BL}
 \bftheta_{t}^*\mid \mathD_{t}, {\bf W}^*_t,\phi &\sim [{\bf m}^*_{t},\phi^{-1}{\bf C}^*_{t}],
  \end{align}
 where $ {\bf m}^*_{t}={\bf a}^*_t+{\bf A}_t(y_t-f_t)$ and ${\bf C}_t^* = {\bf R}^*_t - {\bf A}_tq_t {\bf A}'_t$ and we can easily return to the normality assumption.
\end{myth}

The variance ${\bf W}_t^*$ can be estimated using a discount factor strategy.

\subsubsection{Estimating $\phi$}
\medskip

The steps of the method described above are conditional on $\phi$. In the case where $\phi$ is unknown a practical solution is to use a plug-in estimator for $\phi$ obtained from the maximum a posteriori estimation.

The posterior distribution of $\phi$ given the observed data is given by:
\begin{align}\label{loglik}
 p(\phi \mid \mathD_T) = \prod_{t=1}^{T}  p(y_t\mid\mathD_{t-1}, \phi)p(\phi\mid \mathD_{0}), 
\end{align}
where $p(y_t\mid \mathD_{t-1}, \phi)$ is the predictive distribution conditional on $\phi$. 

Closed-form expressions for the density function of the family of $\cal{NGAL}$ distributions are not available as far as we know. 
However, the density and the cumulative distribution function can be obtained numerically using the convolution form or the inversion of the characteristic function. In particular, using the convolution to represent the 
density function, we get:
\begin{align}\label{apr_1}
p(y_t\mid \mathD_{t-1}, \phi) = \int_{-\infty}^{\infty}{p_{\epsilon}(y_t-z)p_{\zeta}(z)dz},
\end{align}
where $p_{\zeta}(.)$ is the density of the $\cal{GAL}$ distribution, and $p_{\epsilon}(.)$ is the density of a normal distribution with mean $0$ 
and variance $c$. Using the inversion of the characteristic function, we get:
\begin{align}\label{apr_2}
p(y_t\mid \mathD_{t-1}, \phi) = \frac{1}{2\pi}\int_{-\infty}^{\infty}{e^{-isy_t}\varphi(s)ds},
\end{align}
where $\varphi(.)$ is the characteristic function of the $\cal{NGAL}$ distribution.

The integrals (\ref{apr_1}) and (\ref{apr_2}) can be evaluated numerically using current quadrature methods. 
For example, \cite{kuonen2003numerical} discussed numerical integration using Gaussian quadrature and adaptive rules, which dynamically concentrate the computational work in the sub regions where the integrand is most irregular, and the Monte Carlo method. 
They concluded that adaptive Gauss-Kronrod quadrature performed best in their examples.

A brief summary of the algorithm is stated in the following steps: 
\begin{itemize}
 \item [(1)] for $k=0$ give an initial value ${\phi^{-1}}^{(0)};$
\item [(2)] calculate for $t=1,\dots,T$:
\begin{itemize}
\item [(i)] ${\bf a}^*_t={\bf G}^*_t{\bf m}^*_{t-1}$, ${\bf R}^*_t={\bf G}^*_t{\bf C}^*_{t-1}{\bf G}^*_t+{\bf W}_t^*$, $\alpha_t = \phi R_{u,t}^{-1}$ and $\beta_t=\exp(-a_{u,t})\phi R_{u,t}^{-1};$
\item [(ii)] $f_t = {\bf F}_t'{\bf a}_t + a_{\tau}\phi^{-1/2}\alpha_t/\beta_t$ and $q_t=\left({\bf F}_t'{\bf R}_{t} {\bf F}_t + b_{\tau}\alpha_t/\beta_t+a_\tau^2\alpha_t/\beta_t^2\right);$
\item [(iii)] get $p(y_t\mid\mathD_{t-1},\phi^{(k)})$ numerically using (\ref{apr_1}) or (\ref{apr_2});
\item [(iv)] calculate ${\bf m}^*_{t}={\bf a}^*_t+{\bf A}_t(y_t-f_t)$ and ${\bf C}_t^* = {\bf R}^*_t - {\bf A}_tq_t {\bf A}'_t,$ where ${\bf A}_t$ is a function of ${\phi^{-1}}^{(k)}$.
\end{itemize}
\item [(3)] do $k = k+1$ and get ${\phi^{-1}}^{(k)}$ maximizing $p(\phi | \mathD_t)$ in (\ref{loglik}).
\item [(4)] repeat (2) and (3) until convergence is achieved.
\end{itemize}

\section{Applications}

To illustrate the performance of the proposed model and inference procedures, we apply the method to some synthetic data and well-known time series data. 
Then, we apply it to the incidence of tuberculosis in Rio de Janeiro, in which it is important to assess if public health policies are effective, not only in reducing the trend in the number of cases, but also the variability of total cases. 
Moreover, the upper quantiles may be useful to detect an epidemic.

Although the approximated method presents less computing burden and keeps the relevant sequential analysis of the data, 
we interchange the use of the MCMC approach with the approximate method in the following applications.

\subsection {Artificial data examples}

In order to assess the efficiency of the proposed sequential procedure and the convergence of the MCMC estimation, two artificial datasets were generated.
The proposed model was fitted  to these datasets and its estimates were then compared to the true values used in the dataset generation process. 

In both studies a non-informative prior distribution is assumed for the parametric vector with: ${\bf m}_0 = 0$, ${\bf C}_0 = 10^5$, $n_\phi = 0.001$, 
$s_\phi = 0.001$. We take two approaches to deal with the evolution variance: (i) we set an inverse Wishart prior distribution for $\delta=0.95$; and (ii) we apply a discount factor $\delta = 0.95$ setting $W_t = C_t (1 - \delta)/\delta$ \cite[p. 51]{west1997}.

The results shown hereafter for the MCMC algorithm correspond to 5,000 MCMC sweeps, after a burn-in of 1,000 iterations and the chain thinning by taking every 4th sample value.

\subsubsection{Trend and seasonal DLM}\label{sec:trend_sas}
An artificial time series of size $T=100$ was generated for a Gaussian dynamic linear model, with the specification ${\bf F}=(1,0,1,0)'$ and ${\bf G}=\left( \begin{array}{c c} 
 {\bf L_2}& {\bf 0}\\ 
 {\bf 0} & {\bf J_2}(\omega)
  \end{array} \right)$, where ${\bf L}_2=\left( \begin{array}{c c} 
 1 & 1\\ 
 0 & 1\end{array} \right)$ and
 ${\bf J_2}(\omega)=\left( \begin{array}{c c} cos(\omega) &  sin(\omega)\\-sin(\omega) & cos(\omega) \end{array} \right),$
  with $\omega = \frac{2\pi}{12}$.
This corresponds to  a second-order polynomial model with a harmonic component. We arbitrarily fixed $V=49$ and 
${\bf W}=\left( \begin{array}{c c}{\bf W_2} & {\bf 0}\\ 
{\bf 0} & {\bf I_2}\end{array} \right)$, where ${\bf W_2}=\left( \begin{array}{c c}  0.02 & 0.01\\ 
0.01 & 0.01\end{array} \right)$  and ${\bf I_{n\times n}}$ is an identity matrix of dimension $n$.

 The DQLM was fitted for $\tau=0.10,0.50$ and $0.90$ and both inference approaches proposed in the paper were considered. 
 In the MCMC algorithm, we assumed an inverse Wishart prior distribution for ${\bf W}$ with hyperparameters $n_w = 8$ and 
 ${\bf S}_w = 0.1{\bf I}_4$, where ${\bf I}_{4}$. 
 Convergence for all the parameters was achieved. 
 Figures \ref{chain_saz} and \ref{hist_hyperparam} in Appendix C show, respectively, the trace plot with the posterior distribution of 
 parameters $\bftheta_t$'s and the histograms of the posterior densities of some elements of the covariance  matrix ${\bf W}$.
  
\begin{figure}[h!]
\begin{center}
\subfigure[MCMC]{\includegraphics[scale=0.56]{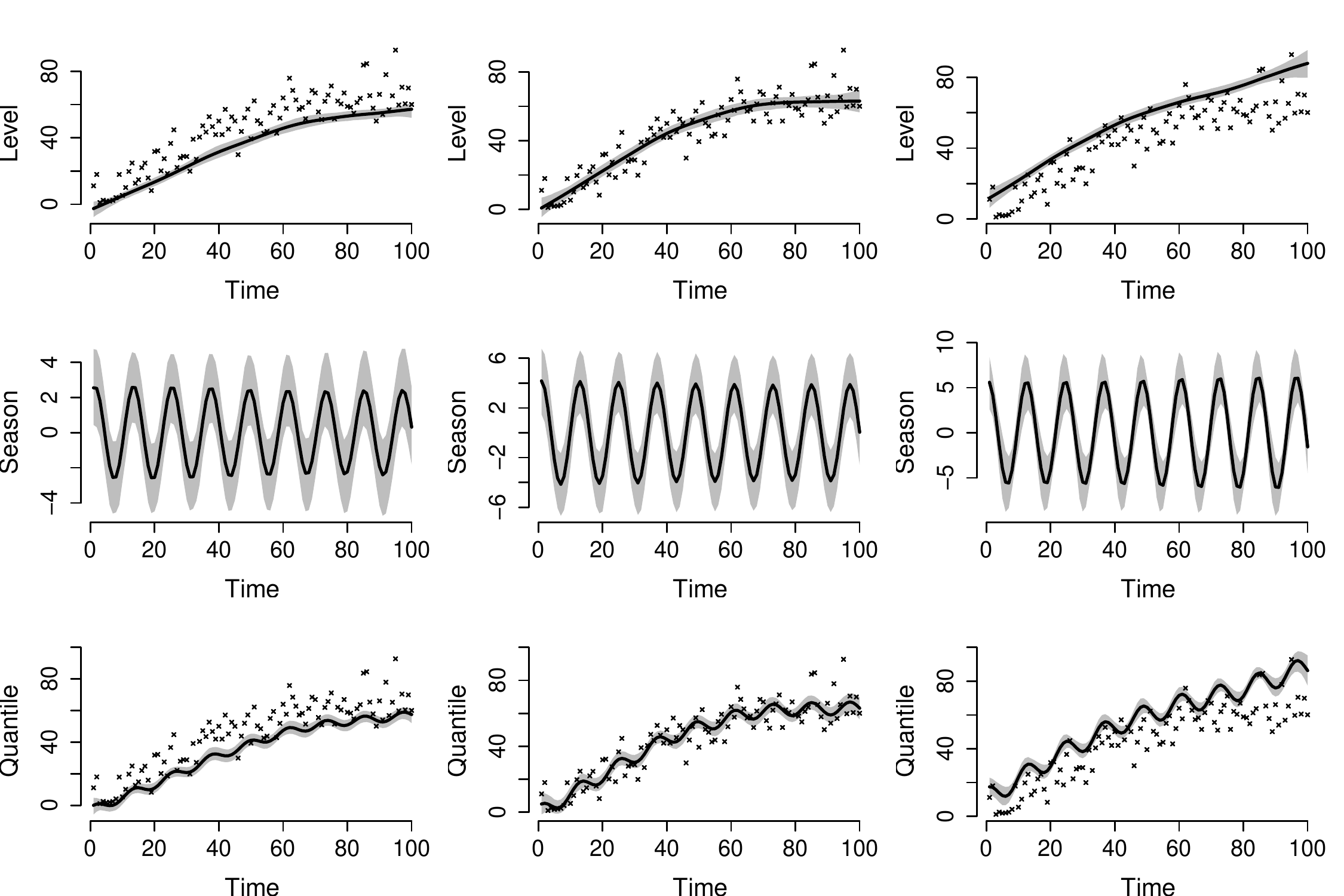}} % salvei como 6 por 9
\subfigure[Approximated method]{\includegraphics[scale=0.56]{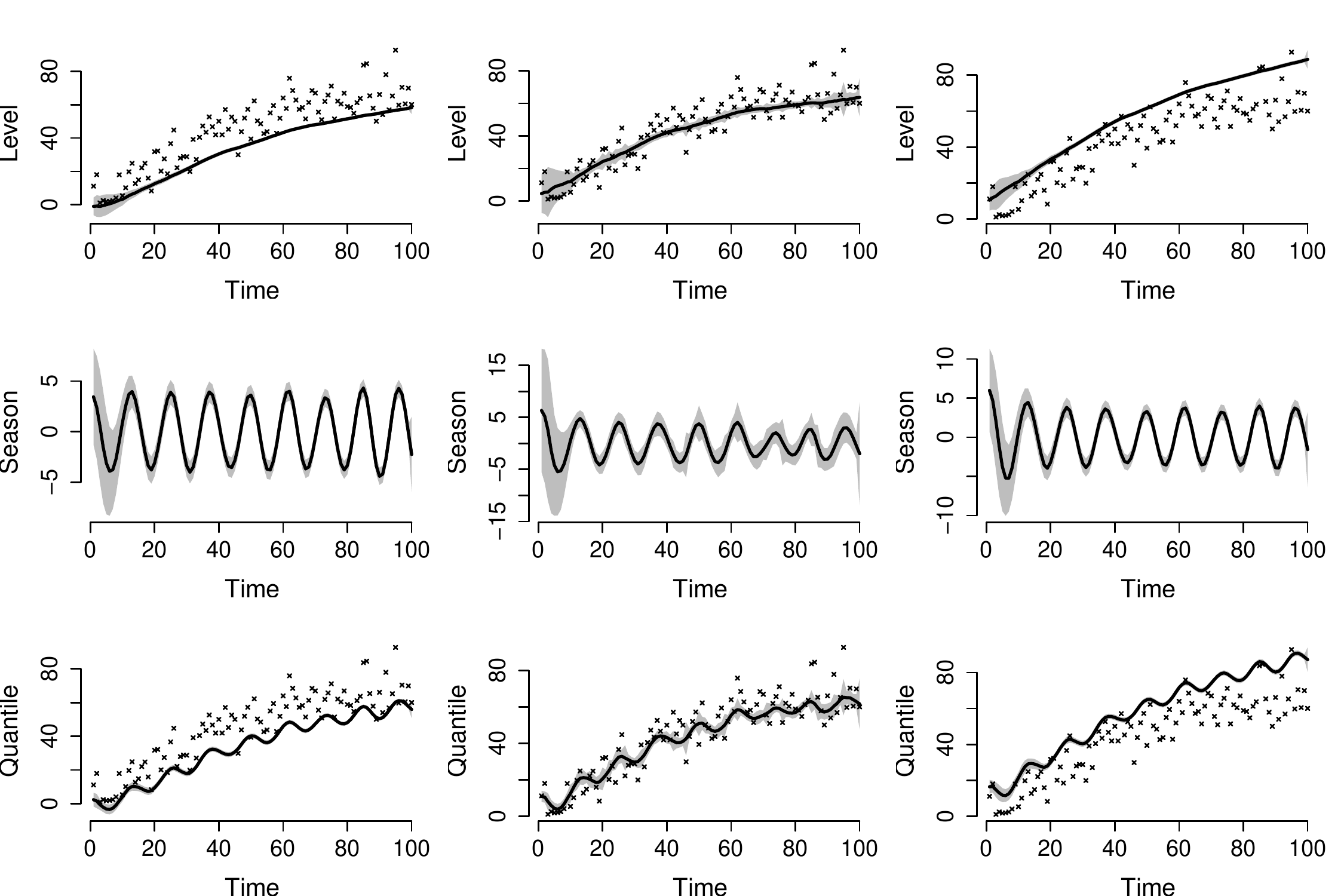}} % salvei como 6 por 9
\end{center}
\vspace{-0.5 cm}\caption{Smoothed posterior mean (solid line), 95\% credible region (shaded area) of the level and seasonal components for each quantile, based on the MCMC output (a) and in the approximated method (b). From left to right $\tau=0.10, 0.50, 0.90$.}\label{thetas}
\end{figure}

Panels of Figure \ref{thetas} present the posterior summary of the level and seasonal components and  
the linear predictor $\bfFt'\bftheta_t$ for each quantile (from left to right $\tau=0.10, 0.50, 0.90$), with the generated  time series  (points). The  posterior mean is represented by the solid line and  
the 95\% credible region by the shaded area. 

\clearpage

 \clearpage

\begin{figure}[h!]
\begin{center}
\hspace{-0.3 cm}\subfigure[10\%]{\includegraphics[scale=0.5]{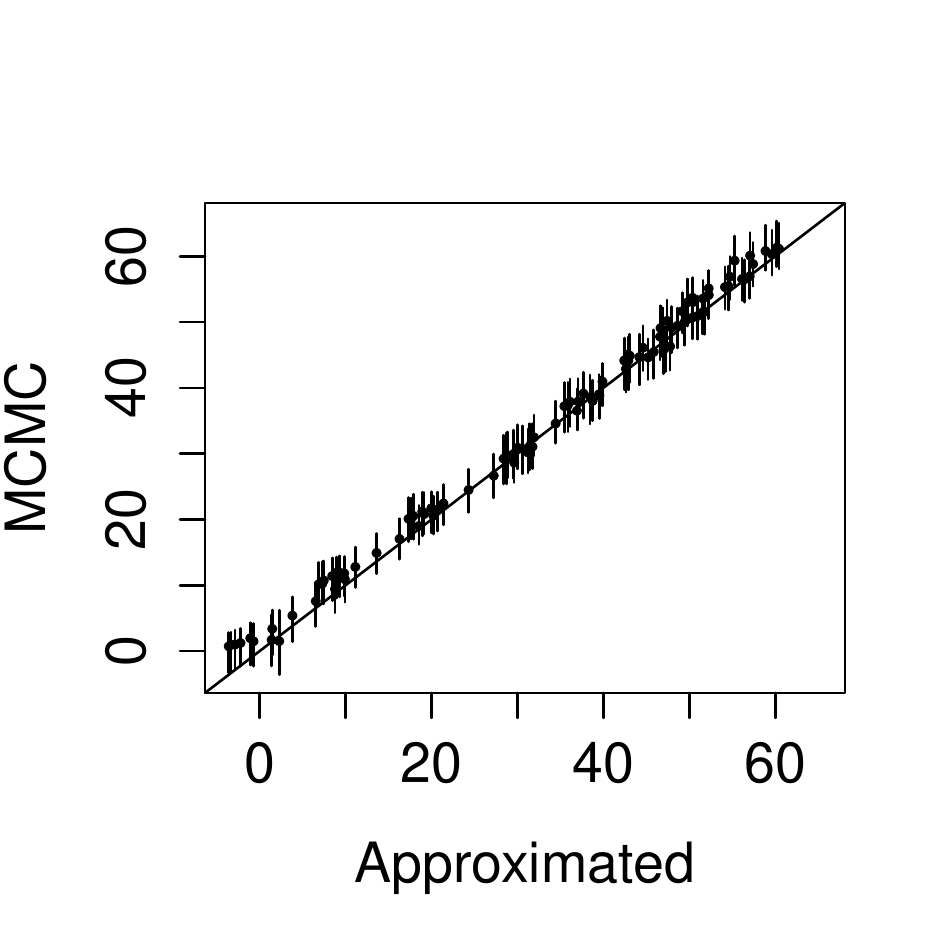}} % salvei como 3.8 por 3.8 no R
\hspace{-0.4 cm}\subfigure[50\%]{\includegraphics[scale=0.5]{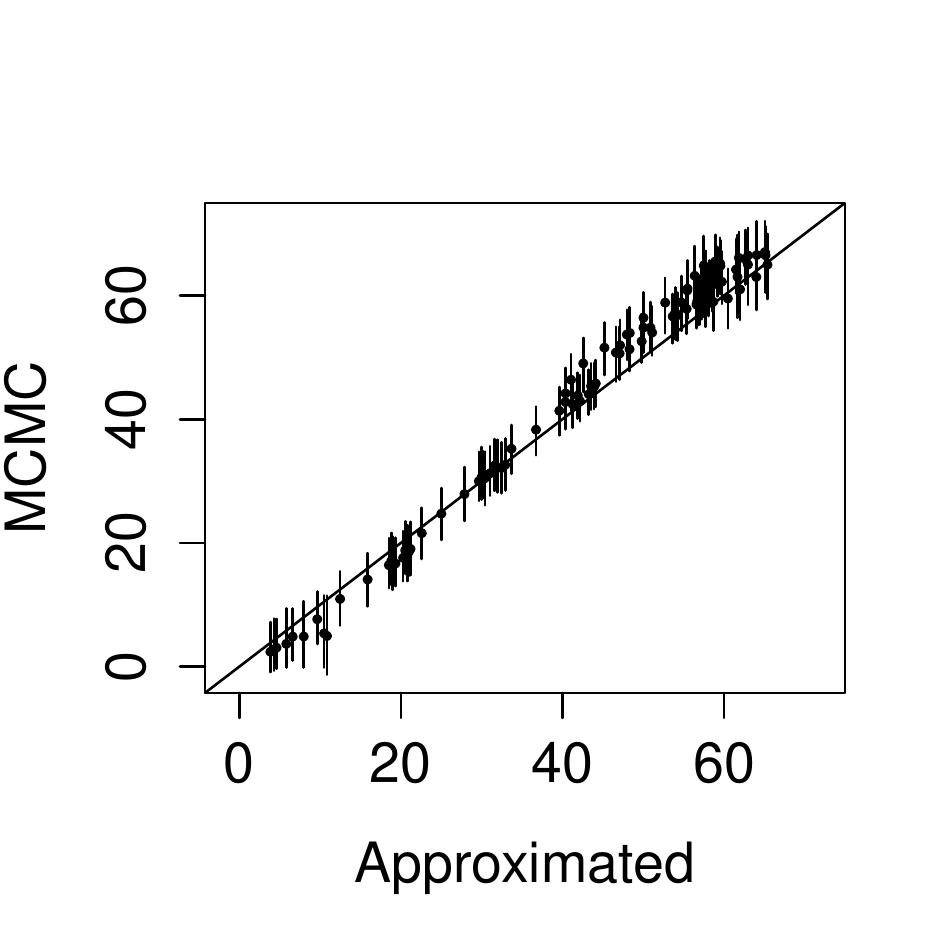}\hspace{-0.4 cm}}
\subfigure[90\%]{\includegraphics[scale=0.5]{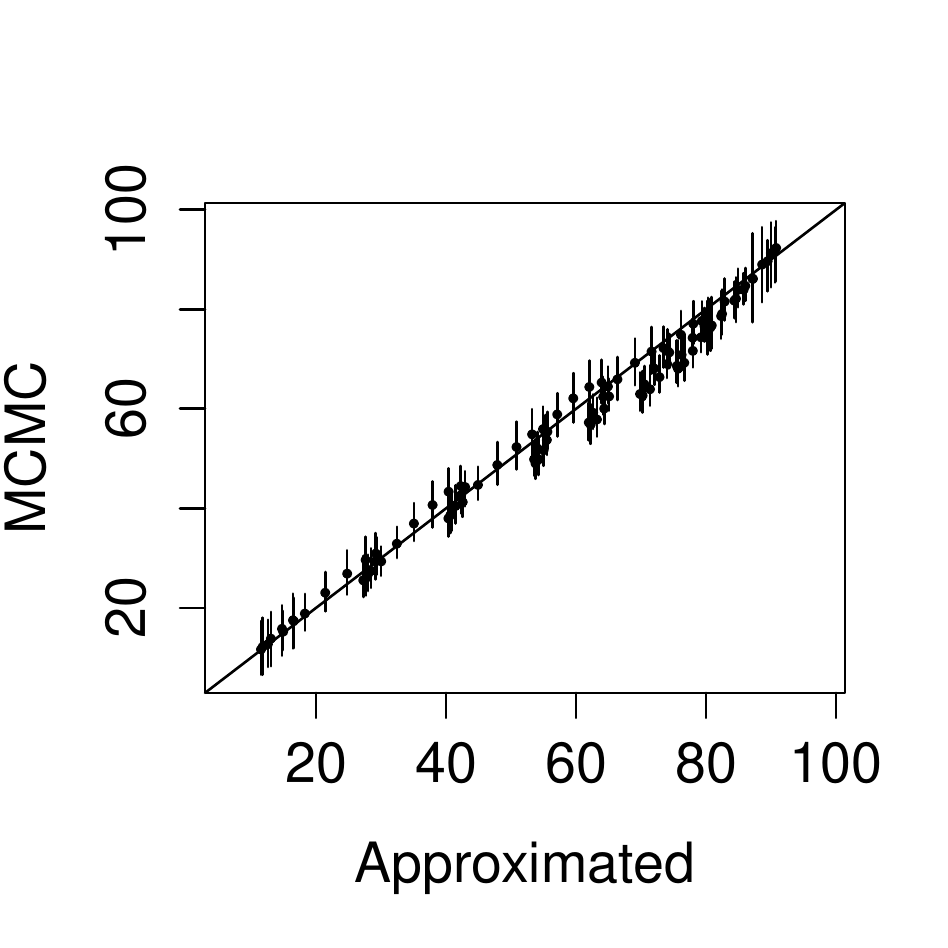}}
\end{center}
\vspace{-0.5 cm}\caption{Plot of the estimated values (posterior mean)  of the linear predictor under the MCMC inference approach versus the
approximated method.}\label{qqplot_seas_trend}
\end{figure}

Figure \ref{qqplot_seas_trend} presents the plot of the estimated values of the linear predictor for each quantile under MCMC approach versus the proposed approximated algorithm. 
The lengths of the segments represent the 95\% credibility interval obtained by MCMC approach. 
We conclude that both methods produce similar results, but while MCMC takes about 5 minutes for 5,000 sweeps for a specific quantile, the approximated method takes seconds. 
Both algorithms were implemented in the R programming language, version 3.4.1 \cite{teamR}, in a computer with an Intel(R) Core(TM) i7-7700 processor 3.60 GHz.
This is first evidence of the relevance of the approximated DQLM  to deal with a scalable dataset.

\subsubsection{Non-Gaussian artificial dataset}\label{exp_fam}

To illustrate how the method works with a non-Gaussian dataset, we generated an artificial dataset from a first order dynamic gamma regression with mean $\mu_t$, scale parameter $\phi$ and a canonical link function $\eta_t = \log(\mu_t)={\bf F}_t'\bftheta_t$. 
In particular, we generated $T=100$ observations assuming ${\bf F}_t=(1,x_{t})$, where $x_t$ is an auxiliary variable at time $t$, 
${\bf G}_t= {\bf I}_{2\times2}$, ${\bf W}=0.01{\bf I}_{2\times2}$, for all $t=1,\dots,T$  and $\phi=50$. 
Each auxiliary variable $x_t$ was generated independently from a uniform distribution defined in the interval (2,4).

The model (\ref{DQLM.2}) was fitted to the original data and to the log transformed data, for the quantiles 0.10, 0.50, and 0.90. We particularly choose here to do the inference using only the MCMC algorithm. Although quantile regression is invariant to monotonic transformations, that is, the quantiles of the transformed variable are the transformed quantiles of the original variable \cite[chapter 2, p. 34]{Koenker2005}, the estimates of all the involved quantities were noticeably better when the data were transformed.

In order to validate the former result, some simulation studies were developed. Several samples were  generated from the gamma model. 
A simple static model is proposed in this exercise,  with ${\bf F}_t=1$ and ${\bf W}_t=0$, for all $t=1,\dots,T$.  
Fifty replications of samples of  sizes $T=100$ and $T=250$ were generated using three  different levels of skewness.

\begin{figure}[h!]
 \begin{center}
 \hspace{-0.3 cm}{\includegraphics[scale=0.78]{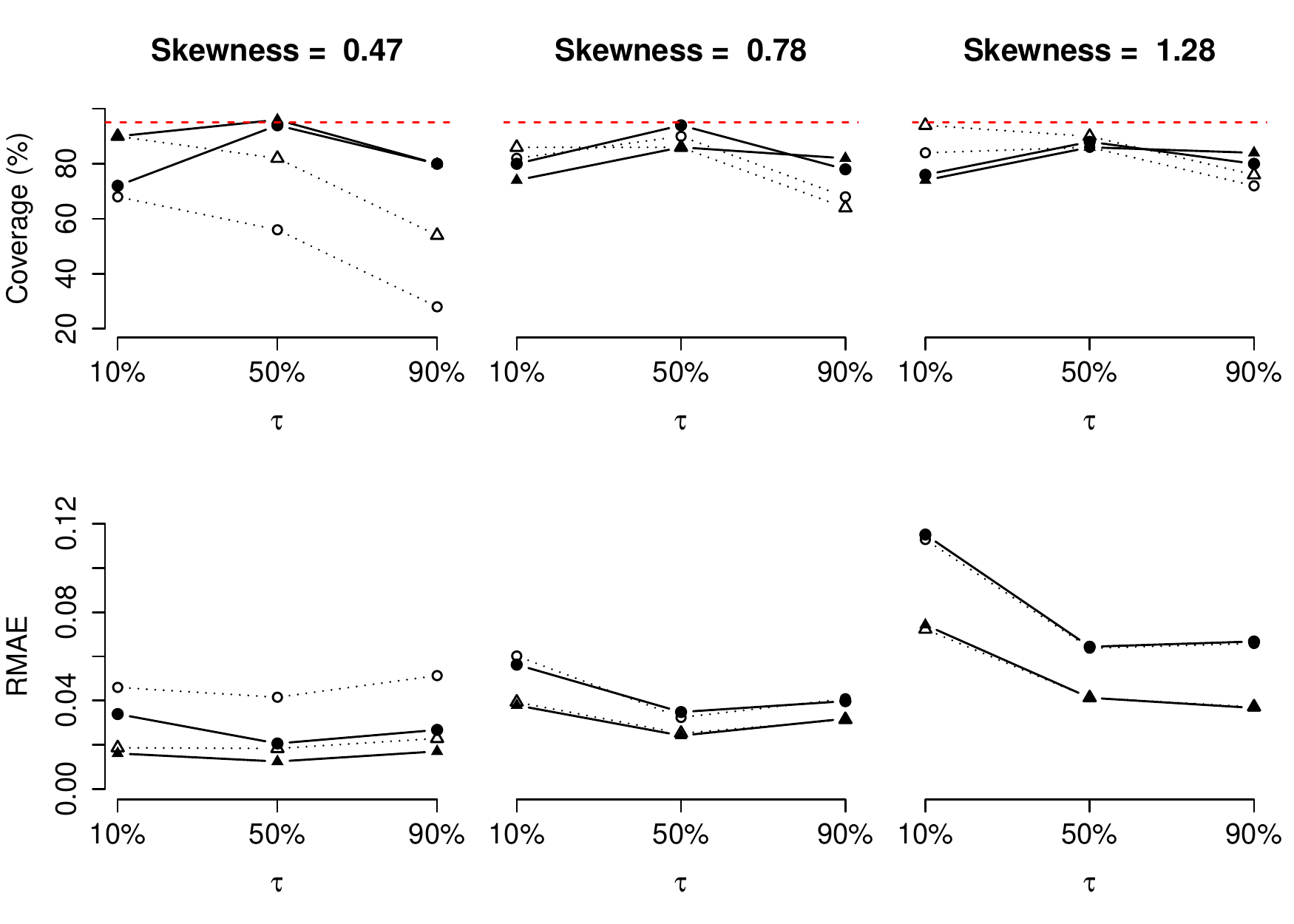}} % salvei como 7 por 5 no R
  \end{center}
 \vspace{-0.7 cm}\caption{Empirical nominal coverage of the 95\% credibility intervals (\%)
and the MAE for the posterior mean of the quantiles for each case. The symbols $\bullet$ and $\blacktriangle$ represent the results obtained when fitting the 
quantile regression to the logarithm of the observations for $T=100$ and $T=250$, respectively. 
$\circ$ and $\vartriangle$ represent the same results when the quantile regression is applied to the observations on the original scale 
for $T=100$ and $T=250$, respectively.}\label{simul_gamma}
 \end{figure}

Figure \ref{simul_gamma} reports the empirical nominal coverage of the 95\% credibility intervals measured in percentages and the relative mean absolute error (RMAE) for the posterior mean of the quantiles for each case.    
The RMAE decreases when we apply the $\cal{AL}$ to the logarithm of the sample observations and also as the sample size increases or the distribution becomes more symmetrically  distributed. The desired nominal level of 95\% is best achieved when the logarithm of the sample is considered, mainly as the level of asymmetry decreases. The improvement in the results when using the log transformation is more noticeable as the skeweness decreases.

\clearpage

Those results encourage us, as a future work, to explore in this context the idea of using the $\mathcal{AL}$ distribution for random
effects in the link function, instead of applying it to the transformed response variable.

\subsection{Real data examples} 

In this section we revisit some classic univariate time series previously analyzed in the literature. In the first example we apply the proposal to the time series of quarterly gas consumption in the UK from 1960 to 1986, in which there is a possible change in the seasonal factor around the third quarter of 1970. 
The second one is the annual flow of the Nile River at Aswan from 1871 to 1970. 
This series shows level shifts, so we considered here a model which includes change points or structural breaks. 
Both datasets are available in the R software \cite{teamR}.

\subsubsection{UK gas consumption}

\begin{figure}[h!]
\begin{center}
\includegraphics[scale=0.5]{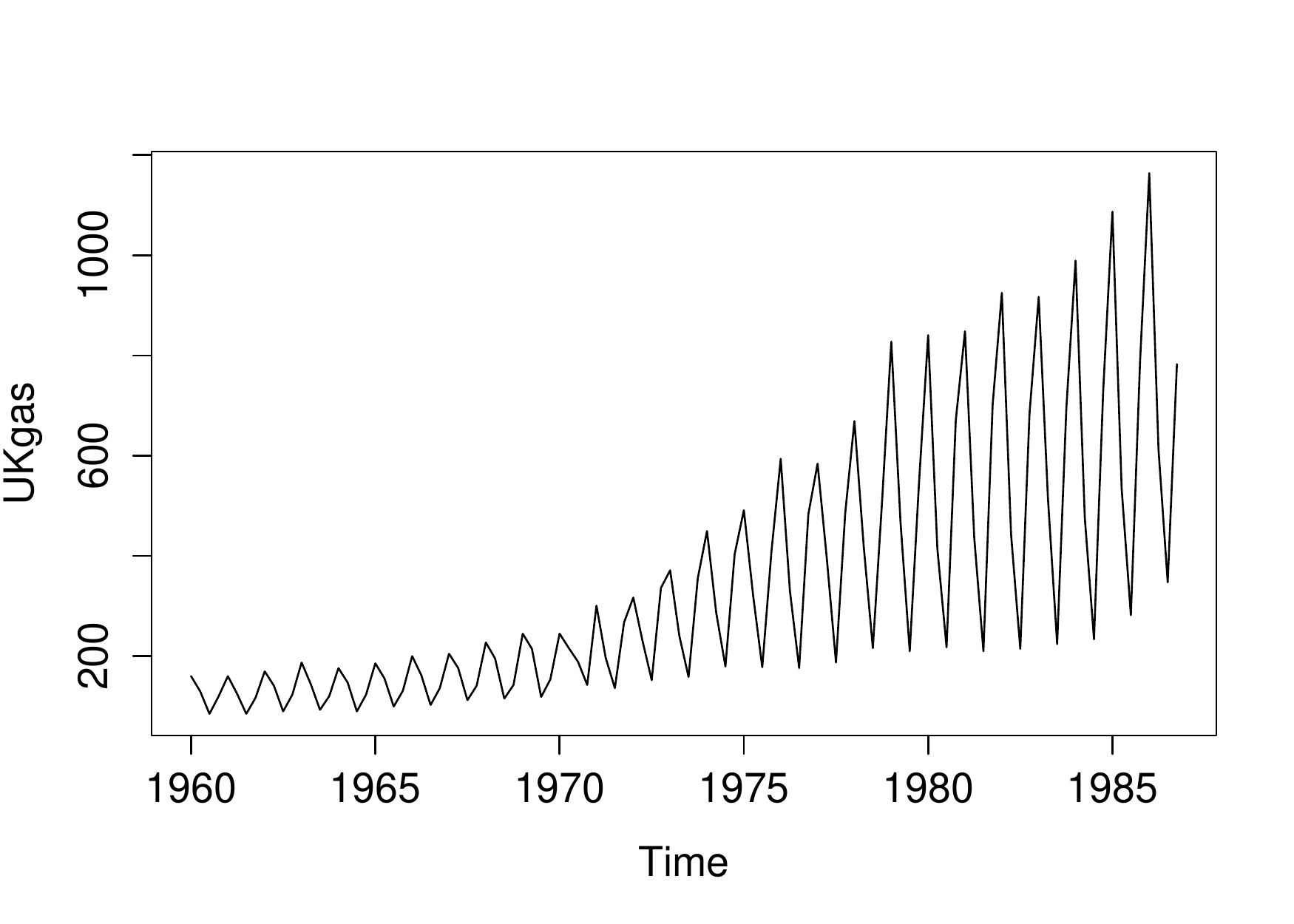}\\ % salvei 7 por 5
\end{center}
\vspace{-1 cm}\caption{Quarterly UK gas consumption from 1960 to 1986, in millions of
therms.}\label{UKgas_data}
\end{figure}

The following dataset consists of quarterly UK gas consumption from 1960 to 1986 (see Figure \ref{UKgas_data}). The plot of the data suggests a
possible change in the seasonal factor around the third quarter of 1970.

We employ a model built on a local linear trend plus a quarterly seasonal component DLM to analyze the data, that is, 
we consider ${\bf F}_t=(1,0,1,0, 0)'$ and ${\bf G}_t= \left( \begin{array}{c c c c c} 
 1 & 1&  0&  0 & 0 \\ 
 0 & 1 & 0 & 0 & 0\\ 
 0 & 0 & -1 & -1 & -1\\
  0 & 0 &  1 & 0 & 0\\  
  0 & 0 &  0 & 1 & 0\\
  \end{array} \right),$ for all $t=1,\dots,108$, in model (\ref{DQLM.2}).
  
\begin{figure}[h!]
\begin{center}
\includegraphics[scale=0.6]{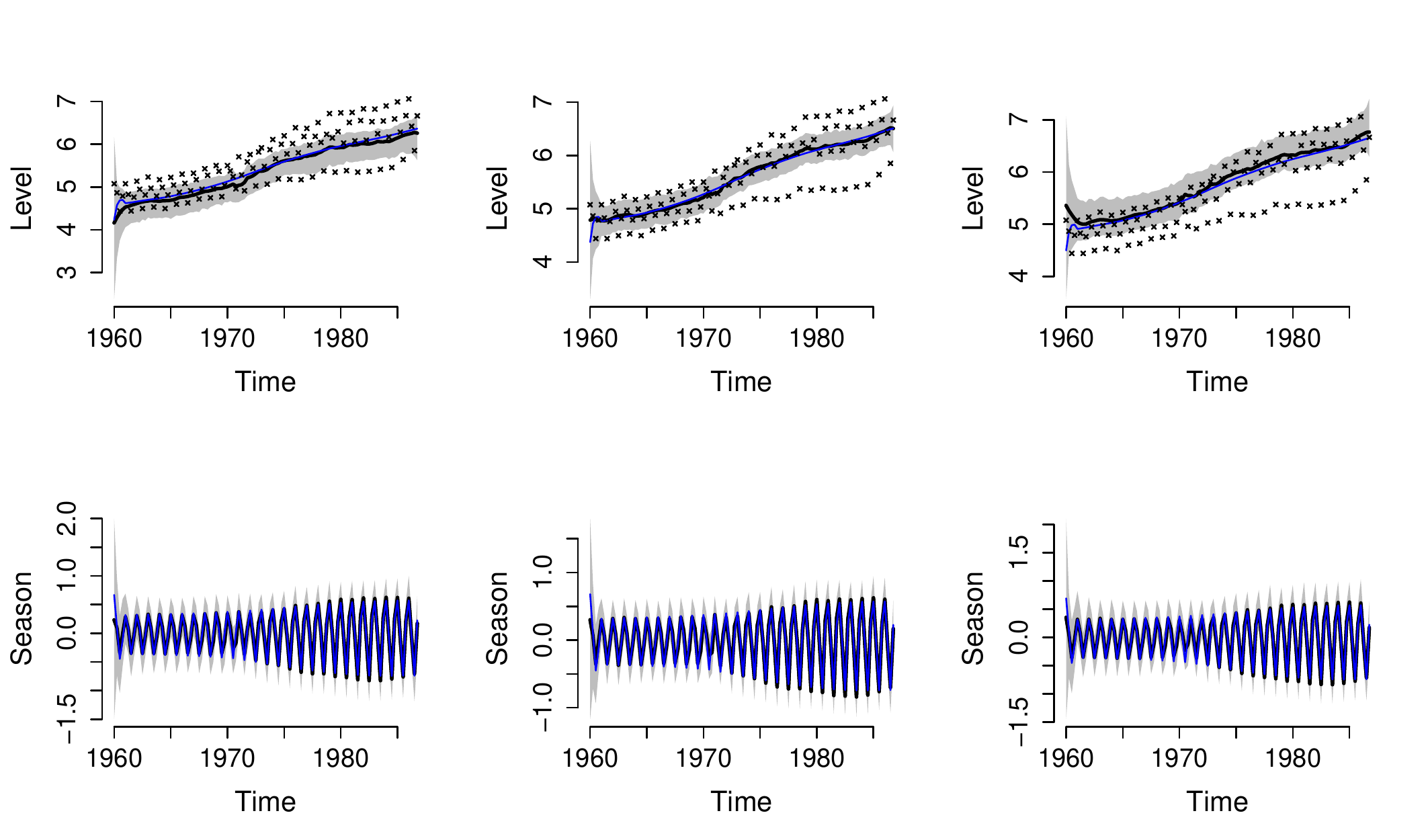} % salvei como 8.5 por 5
\end{center}
\vspace{-0.9 cm}\caption{Posterior mean (solid line), 95\% credible region (shaded area) for the level and seasonal 
components for each quantile.}\label{thetas_UKgas}
\end{figure}

We fit the DQLM for $\tau=0.10, 0.50, 0.90$ to the real dataset on the log scale. Figure \ref{thetas_UKgas} provides a plot of the posterior mean with the MCMC algorithm (in black) and approximated method (in blue) of the trend and seasonal components, together with 95\% credibility intervals 
using MCMC. It is possible to observe the change in the seasonal factor around the third quarter of 1970. Moreover, in terms of the posterior mean 
both methods are very close to each other.
\clearpage

%\clearpage

\subsubsection{Nile River flow}

\begin{figure}[h!]
\begin{center}
\includegraphics[scale=0.4]{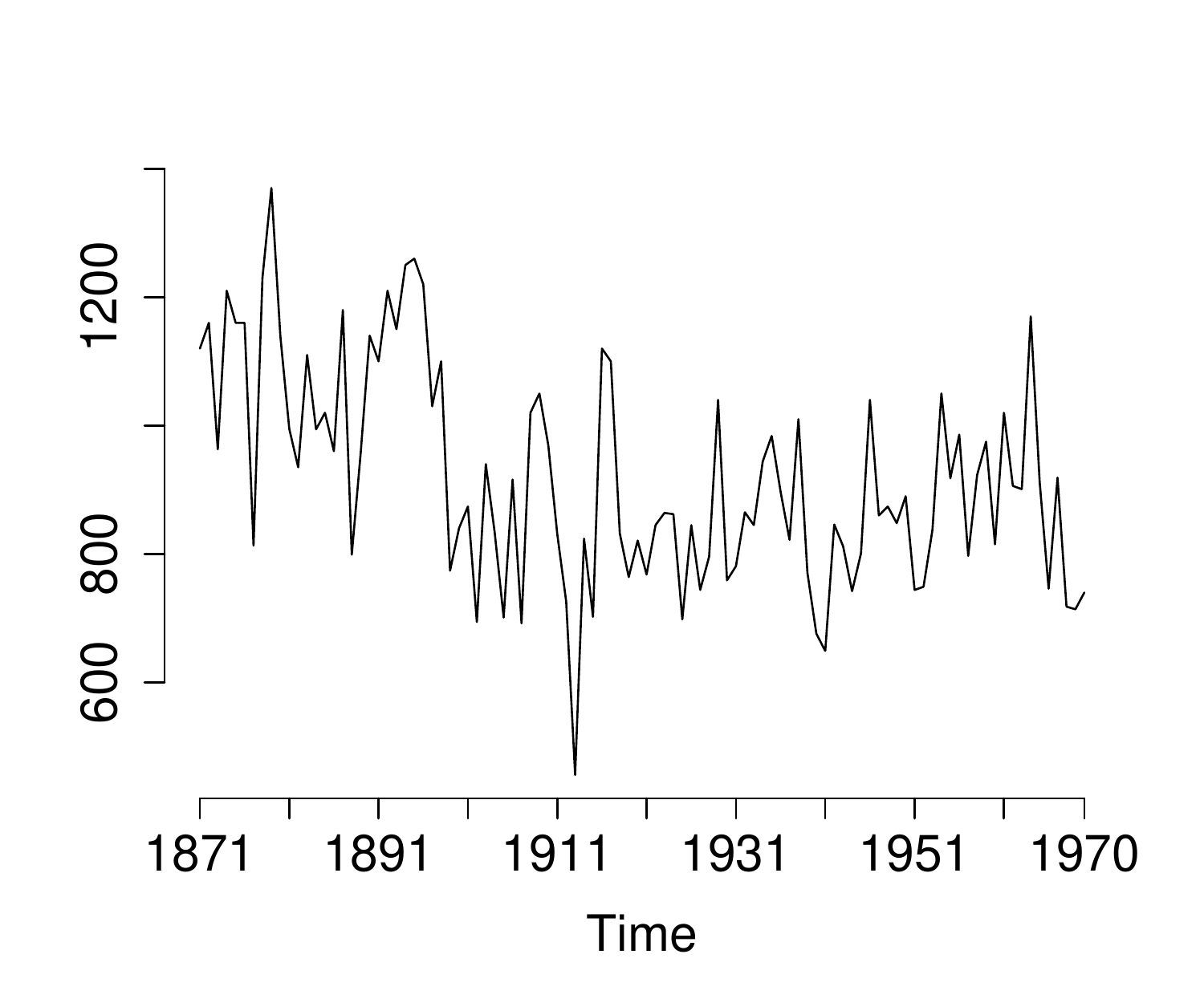}
\end{center}
\vspace{-0.8 cm}\caption{Measurements of the annual flow of Nile River at Aswan from 1871 to 1970.}\label{Nile_data}
\end{figure}

The dataset in Figure \ref{Nile_data} corresponds to the measurements of the annual flow of Nile River at Aswan (Egypt) from 1871 to 1970. 
The time series shows level shifts. 
The construction of the first dam of Aswan started in 1898 and the second big dam was completed in 1971, which caused enormous changes on the Nile flow and in the vast surrounding area. 
Thus, in order to capture these possible level changes we consider here a model that does not assume a regular pattern and stability of the underlying system, but can include change points or structural breaks.

A simple way to account for observations that are unusually far from their one step-ahead predicted value is to describe the evolution error using a heavy-tailed distribution. 
The Student-t distribution family is particularly appealing in this respect for two reasons: 
(i) the Student-t distribution admits a simple representation as a scale mixture of normal distributions, which allows one to treat a DLM with t-distributed observation errors as a Gaussian DLM, conditional on the scale parameters; and
(ii) the FFBS algorithm can still be used. % accommodating through its degrees-of-freedom parameter, different degrees of heaviness in the tails. 
Thus, we consider, the DQLM with evolution characterized by the Student-t distribution, given by:
\begin{align}\label{tst_model}
\begin{array}{rl}
\bftheta_t &\sim \mbox{N}({\bf G}_t\bftheta_{t-1},\lambda_t^{-1}{\bf W}),\\
\lambda_t & \sim \mbox{Ga} (\nu/2,\nu/2), t=1,\dots,T.
\end{array}
\end{align}

The latent variable $\lambda_t^{-1}$ can be informally interpreted as the degree of non-Gaussianity of $w_t$. In fact, values of $\lambda_t^{-1}$ lower than $1$ make larger absolute values of $w_t$ more likely. 
Hence, the posterior distribution of $\lambda_t^{-1}$ can be used to flag possible outliers. 
Through its degree-of-freedom parameter $\nu$, which may also vary on time, different degrees of heaviness in the tails can be attached. This class of models is discussed in \cite{Petris2009dynamic}.

In this example, we fitted a first-order polynomial dynamic model, thus we assumed $F_t=G_t=1$ in model (\ref{DQLM.2}). 
We fitted the model (\ref{tst_model}) and the one with normal evolution, for the 0.25, 0.50, and 0.75-quantiles.

In both cases, for the variance of the states $W$, we considered a half-Cauchy prior distribution, discussed by \cite{gelman2006prior}, with scale 25, set as a weakly informative prior distribution. 
Although we could even assume a prior distribution for this, as described in \cite{Petris2009dynamic}, we assumed 
$\nu$ known a priori and fixed at $2.5$

\begin{figure}[h!]
\begin{center}
{\includegraphics[scale=0.4]{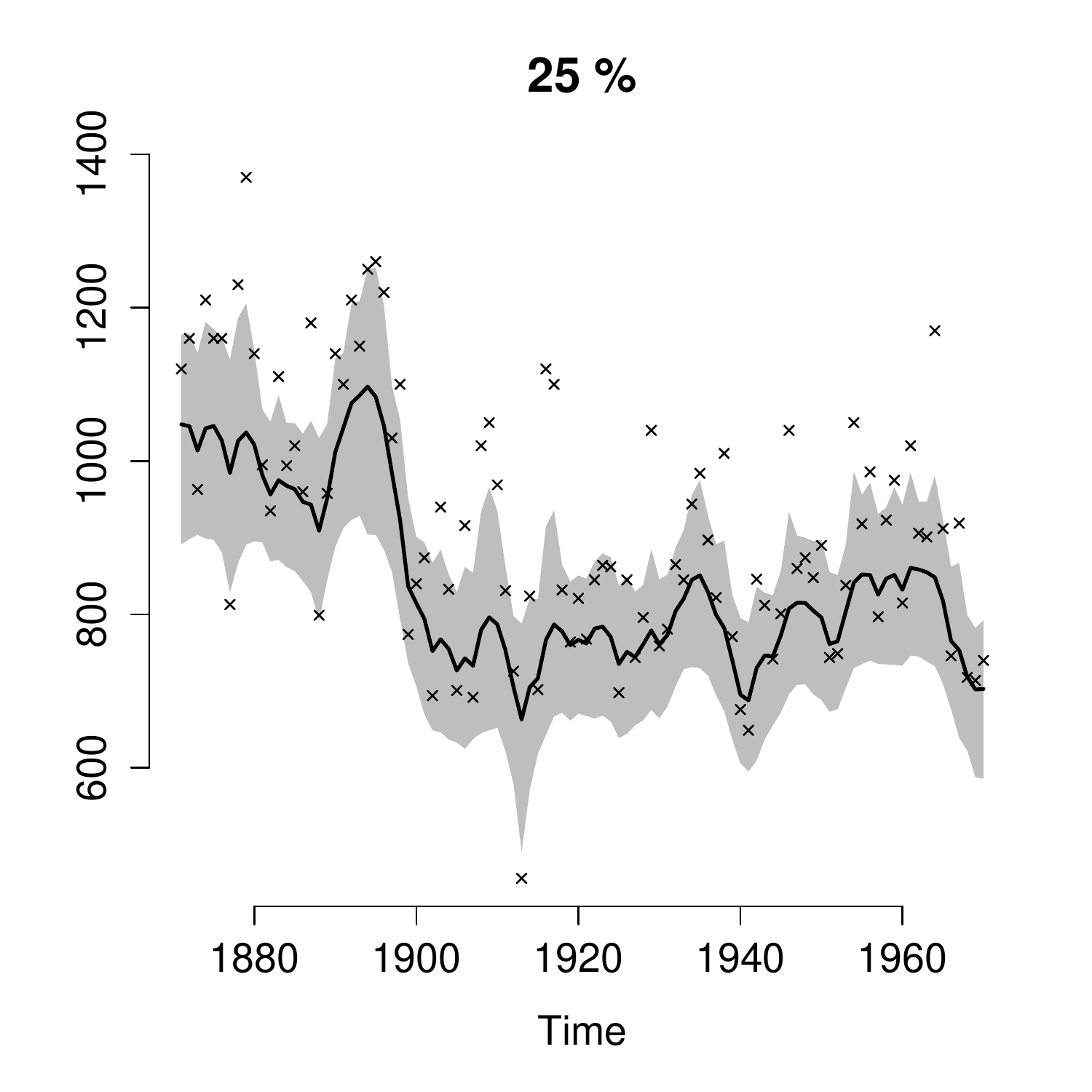}}\hspace{-0.3 cm} % salvei como 6 por 6 no R
{\includegraphics[scale=0.4]{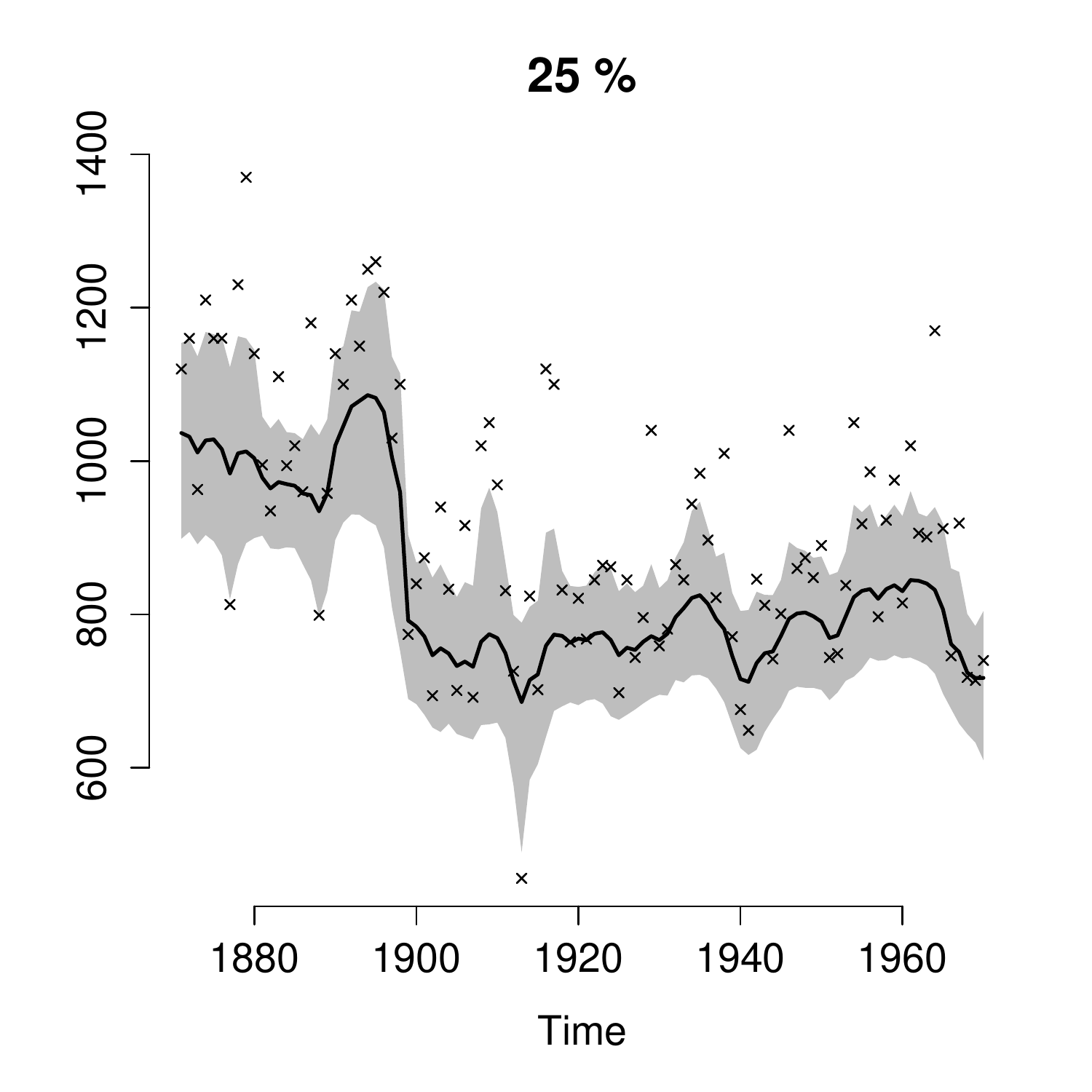}}\\
{\includegraphics[scale=0.4]{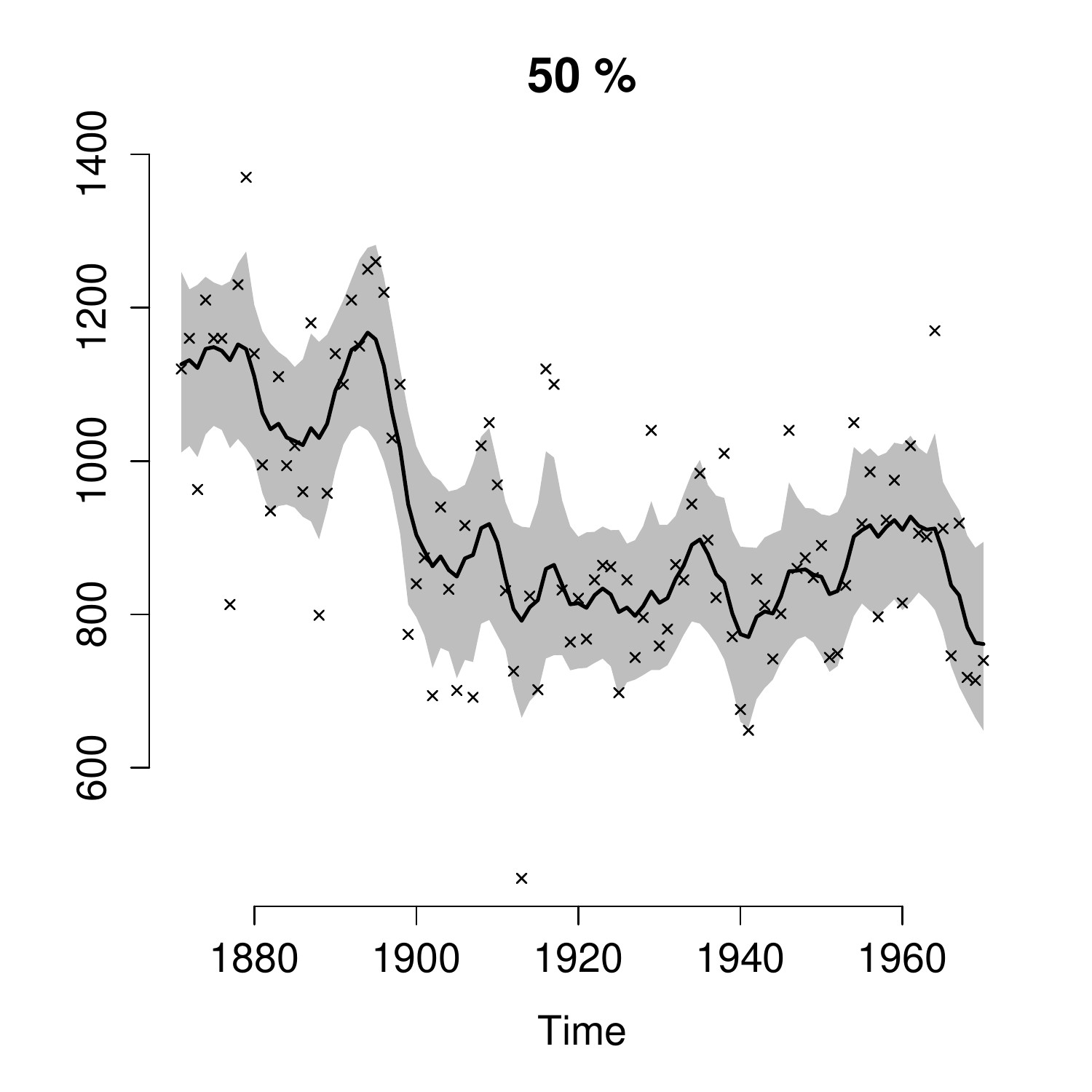}}\hspace{-0.3 cm}
{\includegraphics[scale=0.4]{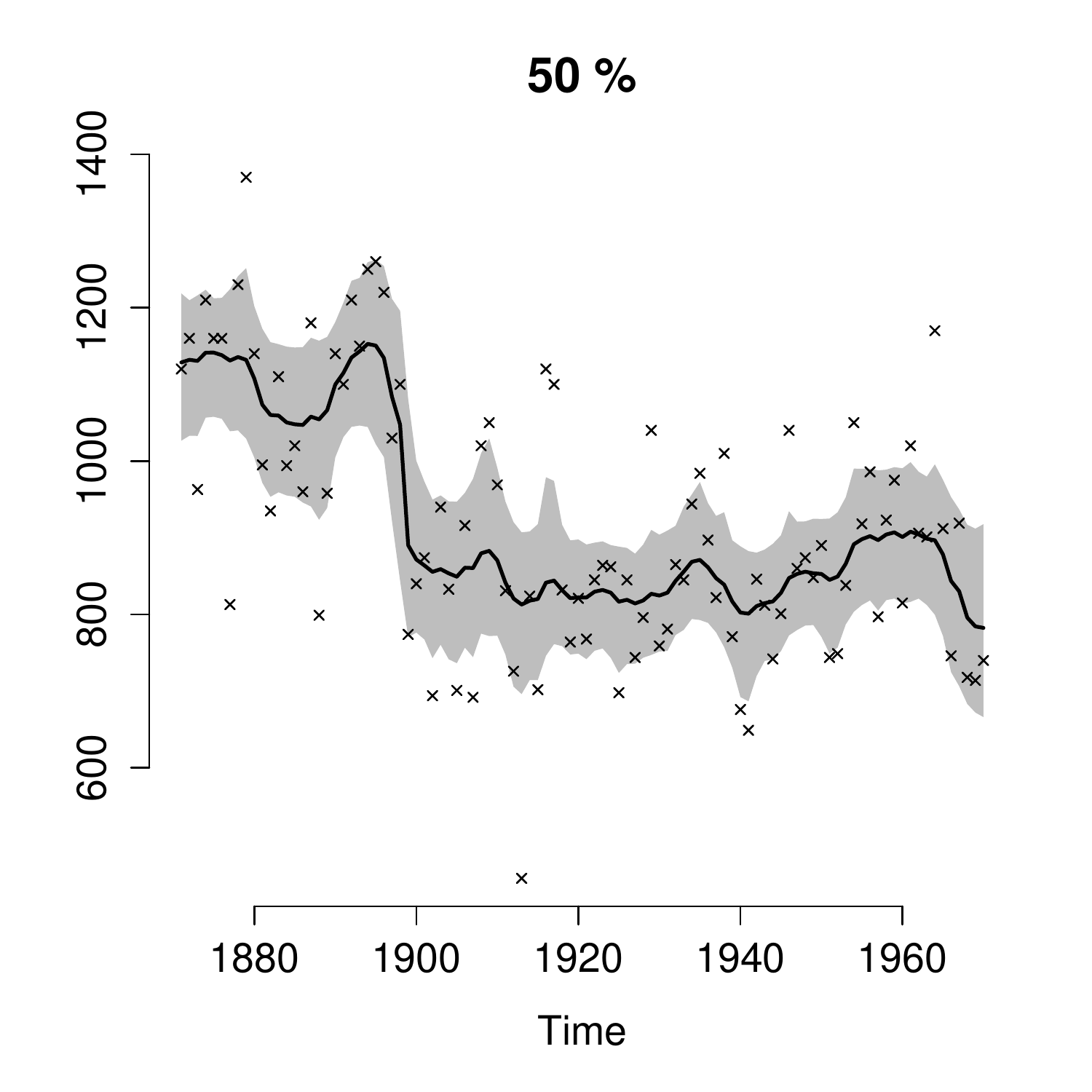}}\\
\subfigure[Normal]{\includegraphics[scale=0.4]{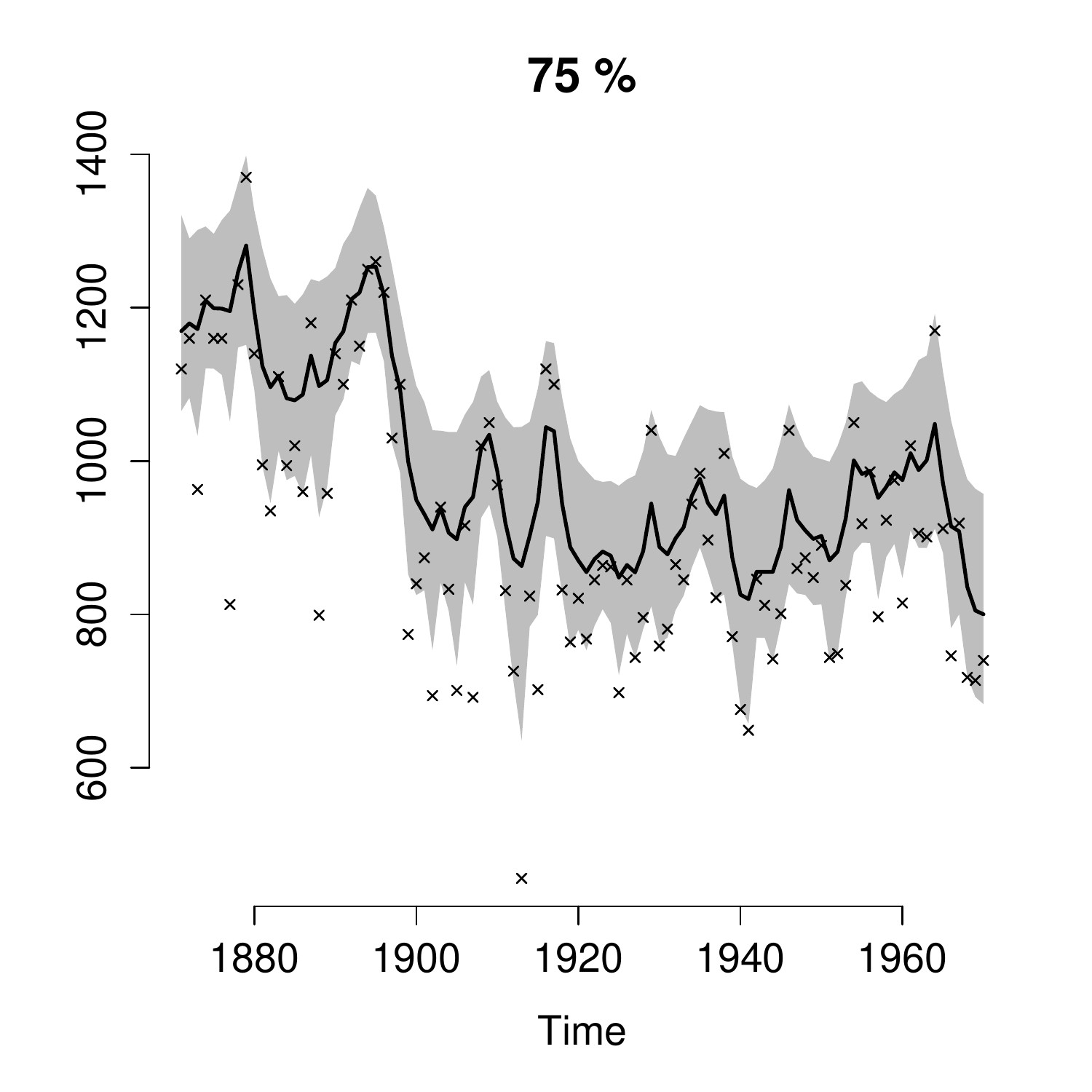}}\hspace{-0.3 cm}
\subfigure[Student-T]{\includegraphics[scale=0.4]{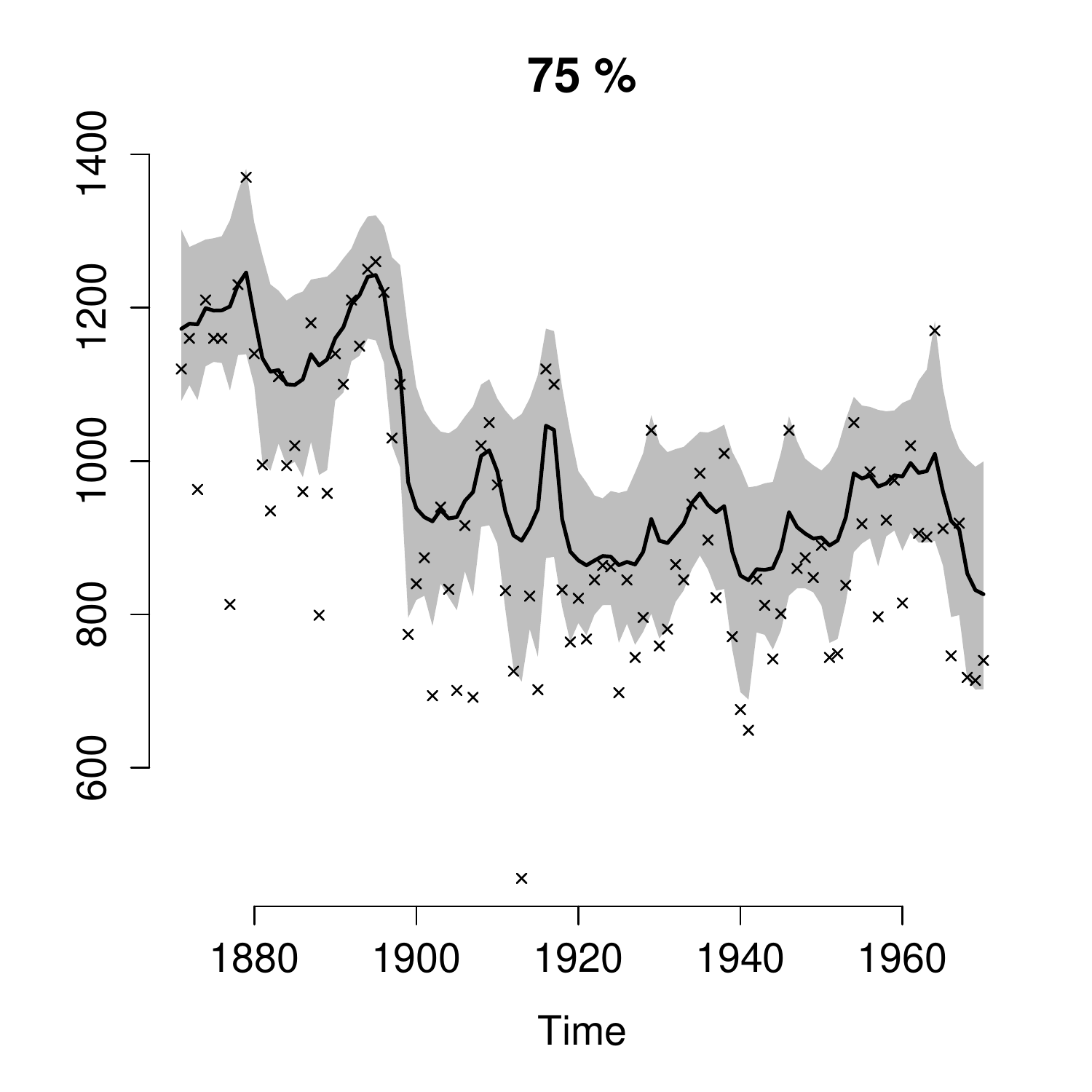}}
\end{center}
\vspace{-0.5 cm}\caption{Posterior mean of the linear predictor (represented by the solid line) and the 95\% credible region (represented by the shaded area) for each quantile.}\label{Nile_normal}
\end{figure}

Figure \ref{Nile_normal} shows the linear predictor for each quantile, with its 95\% credibility interval represented by the shaded area, obtained from the normal (first column) and Student-t fits (second column). 
Model (\ref{tst_model}) results in smoother linear predictors with more accurate credibility intervals.

\clearpage

%\pagebreak
%\newpage

 \clearpage
 
 Table \ref{tabela_Nile} presents the posterior mean of the linear predictor for the 0.10, 0.50, and 0.90-quantiles for the model with normal and Student-t evolution around the year 1899. The abrupt regime change is better captured in the Student-t than in the normal model for all quantiles.

% 
% \pagebreak
% \newpage

\begin{table}[h!]
\caption{Posterior mean of the linear predictor for the 0.10, 0.50, and 0.90-quantiles of 
the model with normal and Student-t evolution.\label{tabela_Nile}}
\centering
\begin{tabular}{|c|ccc|ccc|}
\cline{2-7}
\multicolumn{1}{c}{}&\multicolumn{3}{|c|}{Student-t} &\multicolumn{3}{c|}{Normal}\\\hline
Year & 10\% & 50\% & 90\% & 10\% & 50\% & 90\%\\\hline
%1895 & 1082.35 & 1150.57 & 1242.53 & 1083.60 & 1158.57 & 1253.51\\
1896 & 1064.01 & 1134.51 & 1219.03 & 1046.11 & 1124.00 & 1216.96\\
1897 & 1004.97 & 1083.83 & 1147.69 & 984.23 & 1064.32 & 1137.24\\
1898 &  959.51 & 1047.47 & 1117.77 & 922.98 & 1016.65 & 1093.48\\
1899 & 792.07  & 890.29  & 972.25 & 836.57 & 943.65 & 999.19\\
1900 & 784.17  & 871.67  & 938.27 & 814.69 & 903.88 & 949.23\\
1901 & 771.33  & 863.93  & 926.85 & 794.25 & 882.78 & 930.59\\\hline
%1902 & 746.79  & 855.57  & 921.41 &  752.26 & 862.85 & 910.82\\\hline
\end{tabular}
\end{table}

\begin{figure}[!htb]
\begin{center}
{\includegraphics[scale=0.65]{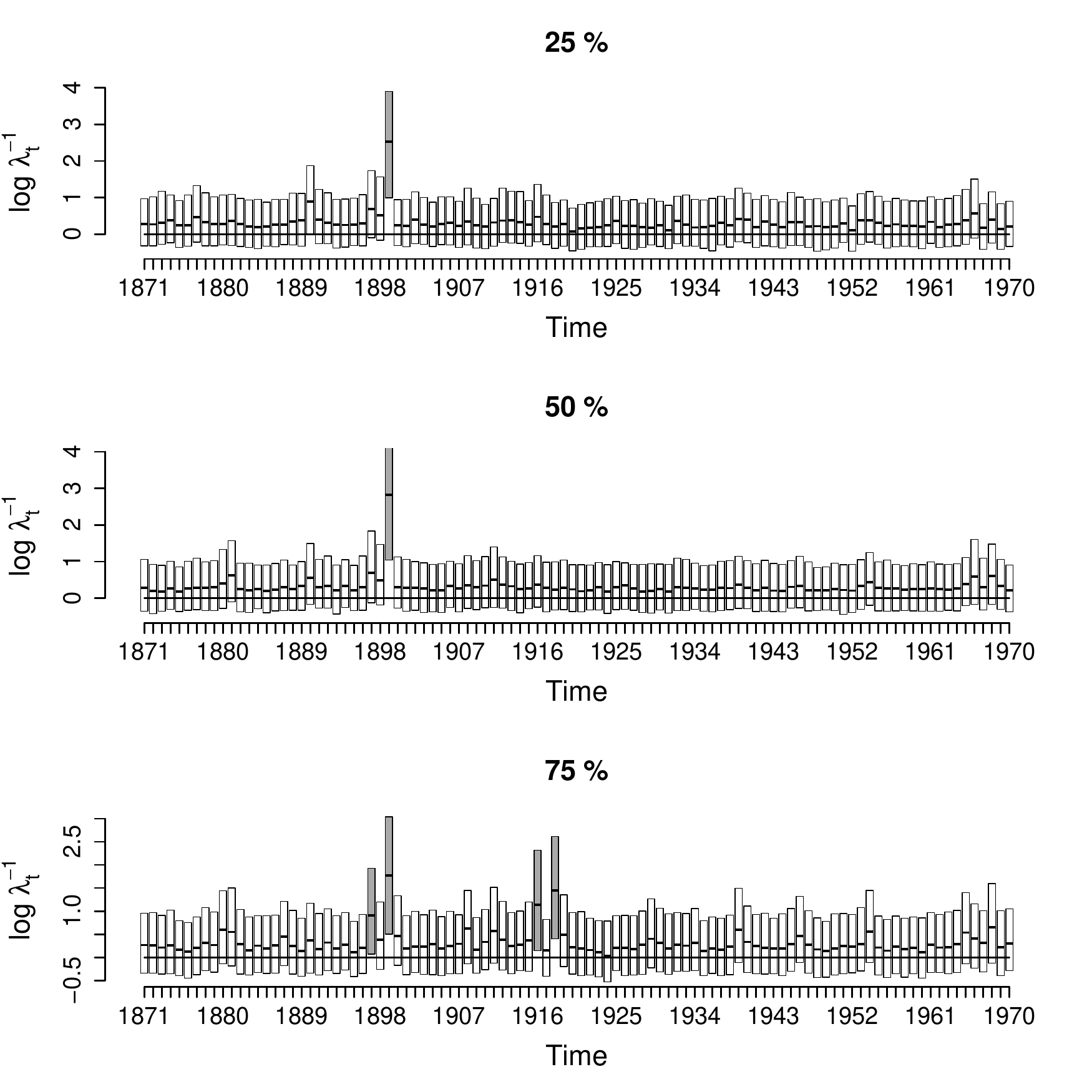}}  
	\vspace{-0.3 cm}\caption{Boxplots of the posterior samples of $\lambda_t^{-1}$ in log scale for the 0.25, 0.5, and 0.75 quantiles over the years.}\label{boxplot_lambdat}
\end{center}	
\end{figure}

Figure \ref{boxplot_lambdat} shows the boxplots of the posterior distribution of $\lambda_t^{-1}$ in logarithmic scale from 1871 to 1970 for each quantile. 
Values of $\log \lambda_t^{-1}$ greater than 0 indicate an abrupt regime change. 
Boxplots in gray do not include the value 0. 
Thus, it is possible to observe that the model in fact accommodates the outliers. 
The regime change in 1899 is detected for all the quantile regression models fitted. 
However, for the 0.75-quantile other outliers were detected.

\clearpage

 \subsection{Tuberculosis cases in Rio de Janeiro}

According to the World Health Organization (WHO), tuberculosis (TB) is one of the top 10 causes of death worldwide. 
Brazil is one of the countries with the highest number of cases in the world and since 2003 the disease has been considered a priority by the Brazilian Ministry of Health. 
As part of the overall effort to reduce the incidence and mortality rate, the Ministry of Health, through the General Office to Coordinate the National Tuberculosis Control Program (CGPNCT), prepared a national plan to end tuberculosis as a public health problem in Brazil, reaching at target of less than 10 cases per 100,000 inhabitants by 2035.
The plan to end the TB epidemic is a target under the Sustainable Development Goals that requires implementing a mix of biomedical, public health and socioeconomic interventions along with research and innovation. 
The World Health Assembly passed a resolution approving with full support the new post-2015 End TB Strategy with its ambitious targets \cite{who2014}. 

\begin{figure}[!hbt]
\begin{center}
\hspace{-0.3 cm}\subfigure[]{\includegraphics[scale=0.42]{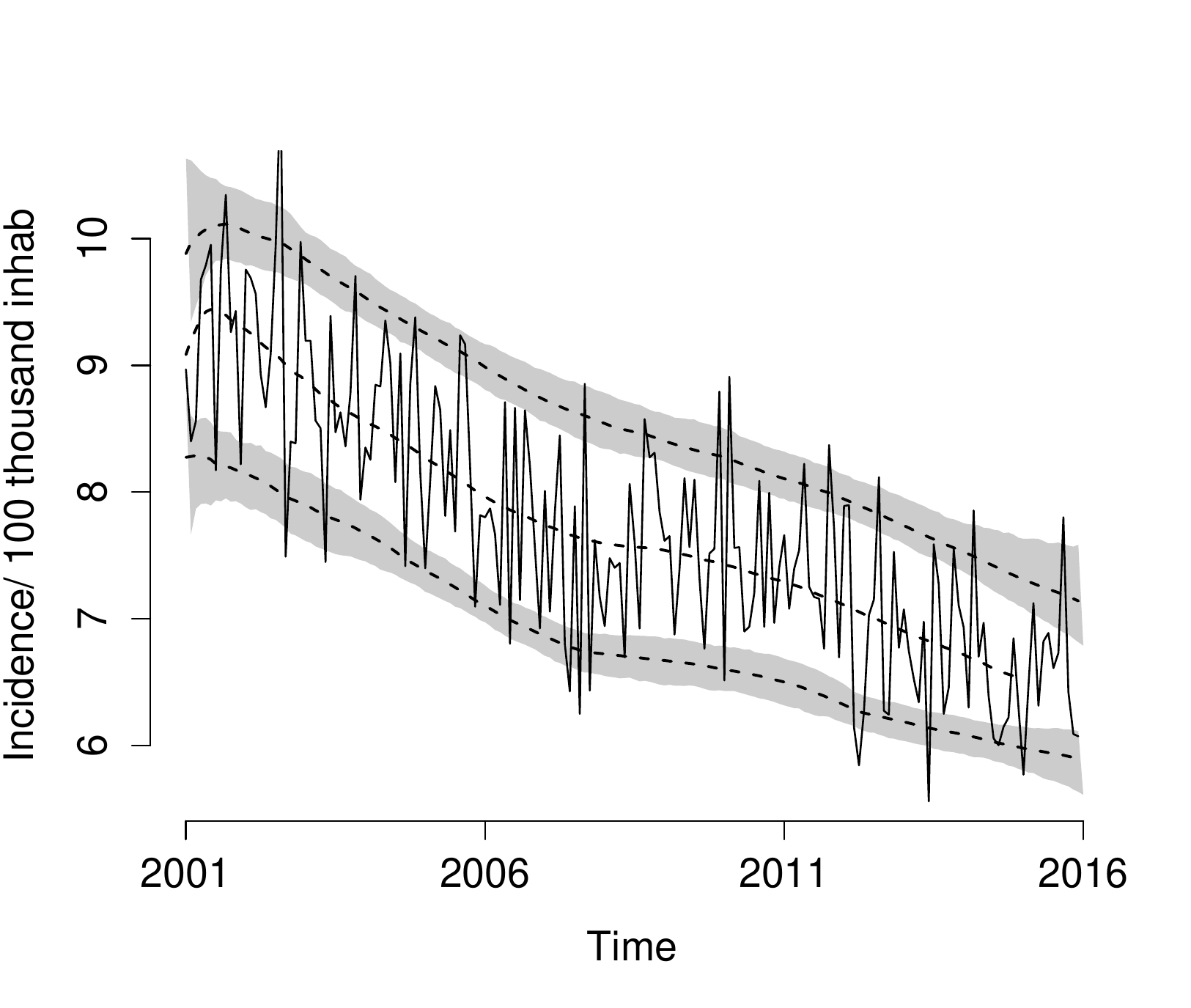}}\hspace{-0.3 cm}
\subfigure[]{\includegraphics[scale=0.42]{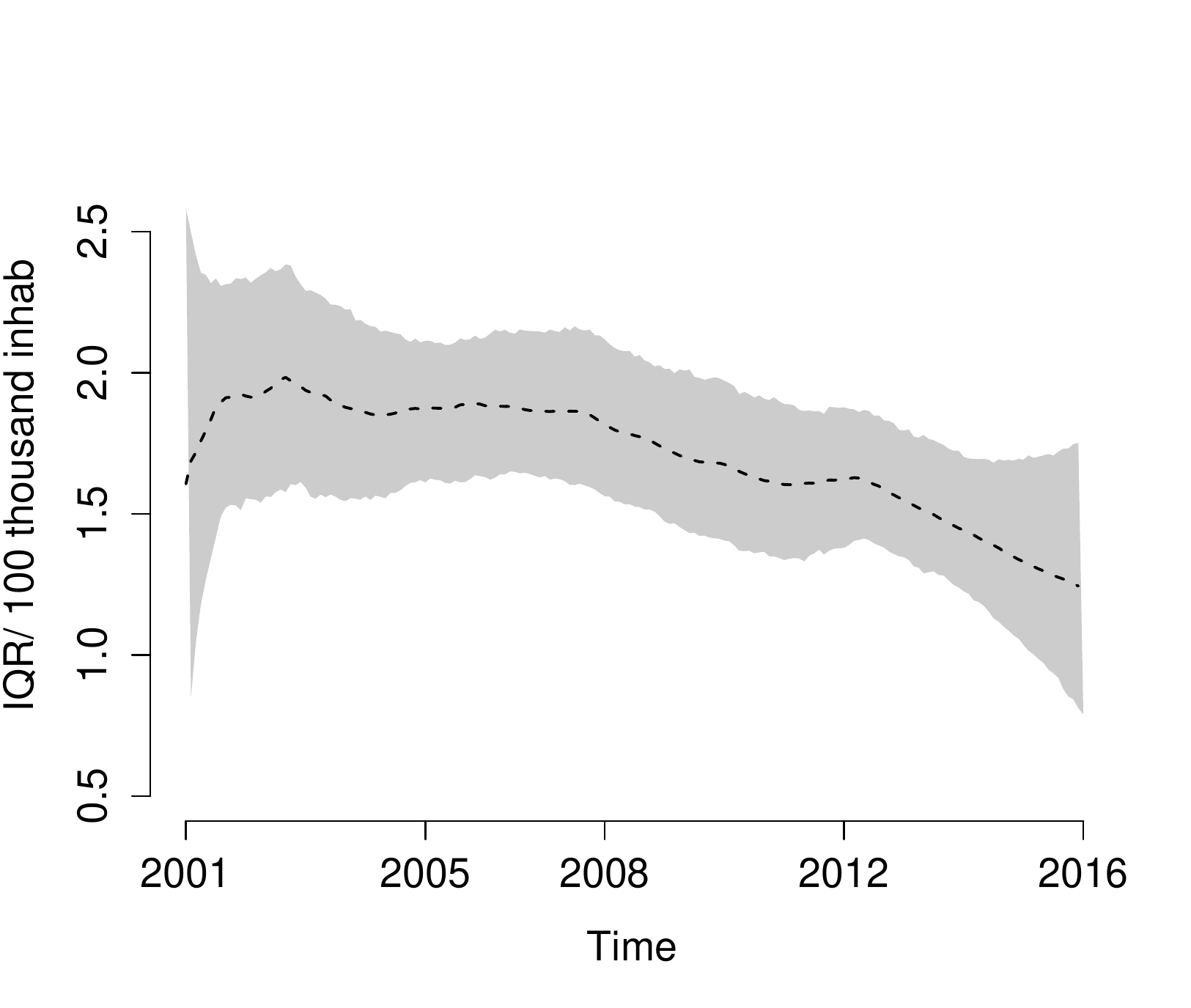}}
\vspace{-0.3 cm}\caption{ Monthly incidence of TB in Rio de Janeiro state: (a) Posterior mean (dashed line) of the linear predictor for the 0.10, 0.50, and 0.90-quantiles, and their 95\% credibility intervals for the 0.10 and 0.90-quantiles; (b) Posterior summary for the interquantile range between the 0.10 and 0.90 quantiles.}
\label{applic_TB_data}
\end{center}
\end{figure}

The state of Rio de Janeiro is located in southeastern Brazil, with over 16 million residents in 2017.
The Rio de Janeiro state has one of the highest TB rates in the country. 
In 2015, there were 13,094  new notified cases, representing 15\% of new cases for the whole country.
We fit the proposed DQLM with a trend component for monthly incidence in Rio de Janeiro from January 2001 to December 2015. 
Figure \ref{applic_TB_data} (a) presents the posterior mean (dashed line) and the 95\% credibility interval (region in gray) for the linear predictor
for the 0.10, 0.50, and 0.90-quantiles, under the MCMC procedure. Figure \ref{applic_TB_data} (b) presents
the posterior summary of the interquantile range (IQR) between 10\% and 90\%. It is possible to observe a decreasing pattern for the IQR, mainly after 2008.

One of the targets of the post-2015 global tuberculosis strategy is a 20\% reduction in tuberculosis incidence by 2020, a 50\% reduction by 2025, an 80\% reduction by 2030 and a 90\% reduction in tuberculosis incidence by 2035, compared with 2015. 
Therefore, one interest here is to predict the incidence for the next 20 years, in order to detect in terms of quantiles, if the target can be achieved. 
For this, we grouped an incidence for all 15 years and predicted the incidence for the next 20 years.

In MCMC inference procedure, samples from ${\bftheta}_{T+k}$, $k$ a non-negative integer, are obtained by propagating the samples from the posterior distribution through the evolution equation (\ref{DQLM.2}). In the approximated method, this is done after estimating $\phi$ through maximum posterior estimation.
Hence, we get that:
\begin{align}\label{post_thet_condU_ksteps}
 \bftheta_{T+k}^*\mid \mathD_{T} &\sim [{\bf a}^*_{T+k},\phi^{-1}{\bf R}^*_{T+k}],
  \end{align}
 where $ {\bf a}^*_{T+k}={\bf G}_{T+k}{\bf a}^*_{T+k-1}$ and ${\bf C}_{T+k}^* = {\bf G}_{T+k}{\bf R}^*_{T+k-1}{\bf G}_{T+k}'$ can be recursively calculated.

\begin{figure}[!hbt]
\begin{center}
\includegraphics[scale=0.6]{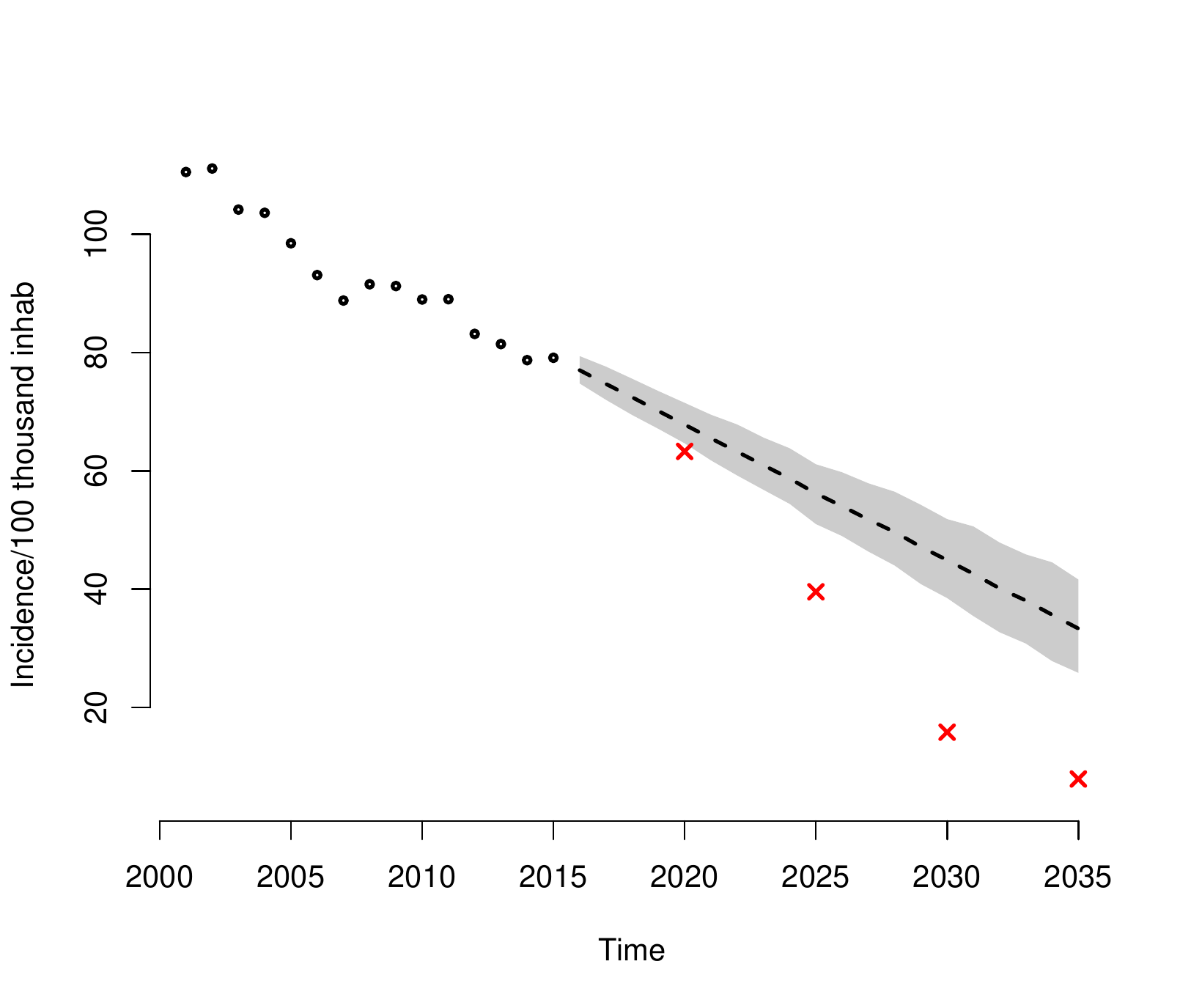} %prev_20anos_TBRJ.pdf}
\vspace{-0.6 cm}\caption{Temporal predictions of annual incidence of TB in Rio de Janeiro state for the next 20 years. The circles represent the observations considered in the inference, the line 
represents the mean of the predictive distribution for the median (0.5-quantile), the region in gray represents the 95\% credibility interval for the median, and the red crosses the reduction targets.}
\label{applic_TB_prev20}
\end{center}
\end{figure}

Figure \ref{applic_TB_prev20} presents the forecast from 2016 to 2035 of the median of the annual incidence of TB per 100 thousand inhabitants under the MCMC approach.
The region in gray represents the 95\% credibility interval and the red crosses indicate the TB reduction targets calculated for Rio de Janeiro state using 2015 as the baseline.
It is possible to observe that the  annual decline in Rio de Janeiro TB incidence rates must accelerate to reach the targets. 

% \pagebreak
% \newpage

% Furthermore, note that the upper quantile overlap the lower one. A possible alternative in this case is to propose a joint model for the quantiles considered or a dynamic quantile hierachical model.

\section{Conclusions} % and suggestions for future work}

In this article we propose a new class of models, named dynamic quantile linear models. 
For the inference procedure, we develop two approaches: 
(i) a MCMC algorithm based on Gibbs sampling and FFBS for the model of the location-scale mixture representation of the asymmetric Laplace distribution; and (ii) a faster sequential method, based on approximations using Kullback-Leibler divergence and Bayes linear method.
The second inference approach has the advantage of being computationally cheaper.

We evaluated the DQLM in artificial and real datasets. 
In the simulation study, we applied our model in a Gaussian example with trend and seasonal components where the DQLM performed well, and the approximate DQLM was a computationally efficient alternative to MCMC.
We also applied our model in a non-Gaussian example generated by a gamma model encouraging the investigation of introducing the $\mathcal{AL}$ model in the link function, or in the response variable.
In the classic real data examples, the UK gas consumption is an example of real data with trend and seasonal components which our model was able to capture. 
The Nile River example, we illustrated by fitting a model for outliers and structural breaks that the detection of occasional abrupt changes differs depending on the quantile of interest.

The application to the tuberculosis data in Rio de Janeiro, Brazil, illustrates the practical importance of evaluating quantiles instead of the mean in the context of forecasting. 
It also encourages us to extend the proposal to joint modeling for the quantiles and a dynamic quantile hierarchical model. Our method can be applied to any infectious disease. It is important to assess whether public health policies are effective, not only in reducing the trend in the number of cases, but also the variability of number of total cases. 
Moreover, the upper quantiles can be useful for early detection of an outbreak (epidemic). 
If the distribution of the number of cases (represented by the quantiles) is much higher than usual, it is a strong indication that attention is required.

%\clearpage

\section*{Appendix A: Proof of Lemma \ref{best_lemma}}% and Theorem \ref{LB}}

\subsection*{Lemma \ref{best_lemma}}
(i) Let us find a tractable monotone transformation $\zeta = \zeta(\theta)$ to induce normality in the sense of minimizing the divergence measure between $p(\zeta)$ and its normal approximation. In particular we have $\theta\sim Ga(a,b)$. This is equivalent to finding $\zeta$ that minimizes the expected 
value
\begin{align*}
k[p(\zeta),q(\zeta)] = \int{p(\zeta)\log{\frac{p(\zeta)}{q(\zeta)} d\zeta}}, 
\end{align*}
where $q(\zeta)$ is the density function of the $N[E(\zeta), V(\zeta)]$ distribution and $p(\zeta)=p(\theta)/|d\zeta/d\theta|.$ 

Thus, we want a function $\zeta$ which minimizes
\begin{align*}
k[p(\zeta),q(\zeta)] &= \int{p(\zeta)\log{p(\zeta)}d\zeta}-\int{p(\zeta)\log{q(\zeta)}d\zeta}\\
&= \int{p(\zeta)\log{p(\zeta)} d\zeta}+\frac{1}{2}\log(2\pi V(\zeta))\int{p(\zeta)d\zeta}+\int{p(\zeta)\left[\frac{(\zeta-E(\zeta))^2}{2V(\zeta)}\right]d\zeta}\\
&= \int{p(\zeta)\log{p(\zeta)} d\zeta}+\frac{1}{2}\log(2\pi V(\zeta)). 
\end{align*}

We consider the class of transformations $\zeta=\zeta(\theta)$, such that $d\zeta/d\theta=\theta^{-\alpha}$, $0\leq\alpha\leq 1$, which contains as 
particular cases 
the standard transformations:
\begin{align*}
\zeta (\theta) &=\left\{\begin{array}{lr} \theta, & \mbox{ for } \alpha = 0,\\
\log\theta, & \mbox{ for } \alpha = 1,\\
(1-\alpha)\theta^{1-\alpha}, &\mbox{ for } \alpha < 1.
\end{array}
\right.
\end{align*}

We have
\begin{align*}
\int{p(\zeta)\log{p(\zeta)} d\zeta} &%= \int{p(\theta)\left[\log{p(\theta)}-\log{\zeta'(\theta)}\right] d\theta} 
= \int{p(\theta)\log{p(\theta)}d\theta}-\int{\log{\zeta'(\theta)}p(\theta)d\theta}\\
& = C_1 + \alpha\int{\log \theta p(\theta) d\theta} = C_1+ \alpha E(\log \theta) \approx C + \alpha\left(\log \frac{a}{b}-\frac{1}{2a}\right).
\end{align*}

Moreover, using that $V(\zeta) \approx \left[\zeta'(E(\theta))\right]^2 V(\theta)$, we get $\log(2\pi V(\zeta)) = 2\alpha \log\left(b/a\right)+C_2.$ Then,
%$V(\zeta) = \left(\frac{b}{a}\right)^{2\alpha}\frac{a}{b^2}$ and
%\begin{align*}
%%\log(2\pi V(\zeta)) = \log(2\pi)+\log\left(\frac{a}{b^2}\right)+2\alpha \log\left(\frac{b}{a}\right).
%\log(2\pi V(\zeta)) = 2\alpha \log\left(b/a\right)+C_2. 
%\end{align*}
\begin{align*}
k[p,q] %& = \alpha\log \frac{a}{b} - 2\alpha \log \frac{a}{b}-\alpha \frac{1}{2a} + C, \\  
& = \alpha\left(\log \frac{b}{a} -\frac{1}{2a}\right)+C,
%& = a\log b - \log \Gamma(a) + (a-1)\int{\log\theta\, p(\theta) d\theta} +(\alpha-b)\int{\theta \, p(\theta) d\theta}\\
%& = a\log b - \log \Gamma(a) + (a-1)E\left(\log\theta\right)+(\alpha-b)E\left(\theta\right)+\frac{1}{2}\log(2\pi V(\zeta))\\
%& = a\log b - \log \Gamma(a) + (a-1)\left[\log a -\psi(b+1)\right] +(\alpha-b)\frac{a}{b}+\frac{1}{2}\log(2\pi V(\zeta)).
%& = a\log \left(\frac{b}{a}\right)-\log \Gamma(a+1) - (a-1)\psi(b+1) - a + \alpha \frac{a}{b} + \frac{1}{2}\log(2\pi)+\frac{1}{2}\log\left(\frac{a}{b^2}\right)+\alpha \log\left(\frac{b}{a}\right)
\end{align*}
%Using that $V(\zeta) \approx \left[f'(E(\theta))\right]^2 V(\theta)$, we get $V(\zeta) = \left(\frac{b}{a}\right)^{2\alpha}\frac{a}{b^2}$ and
%\begin{align*}
%\log(2\pi V(\zeta)) = \log(2\pi)+\log\left(\frac{a}{b^2}\right)+2\alpha \log\left(\frac{b}{a}\right). 
%\end{align*}
which is a decreasing function in $\alpha$ for the particular values that $a$ and $b$ can assume in the method. It follows that progressively better normalizing transformations are obtained for larger values of $\alpha$. 
Thus, in this class of transformations, $\zeta(\theta)=\log\theta$ is the one which minimizes the divergence measure between $p(\zeta)$ and its normal approximation.\\

(ii) If we have $\zeta\sim N[\mu,\sigma^2]$, such that $\mu=E(\zeta)$ and $\sigma^2=V(\zeta)$, then $\theta=\exp(\zeta)$ has distribution 
$LN[\mu,\sigma^2]$ and we desire the better $Ga(a,b)$ to approximate this lognormal distribution in the sense of Kullback-Leibler divergence. That is, we want $a$ and $b$ which minimize
\begin{align*}
k[p(\theta),q(\theta)] &= \int{p(\theta)\log{p(\theta)}d\theta}-\int{p(\theta)\log{q(\theta)}d\theta},
\end{align*}
or maximizes
\begin{align*}
f(a,b)=\int{p(\theta)\log{q(\theta)}d\theta},
\end{align*}
where $q(\theta)$ and $p(\theta)$ are the density functions of the $Ga(a,b)$ and $LN(\mu,\sigma^2)$ distribution, respectively.

Thus,
\begin{align*}
f(a,b) = \int{p(\theta)\log{q(\theta)}d\theta} &= \int{\left(a\log b - \log(\Gamma(a))+(a-1)\log(\theta)-b\theta\right)p(\theta)d\theta}\\
& = a\log b - \log(\Gamma(a)) + (a-1)E_{LN}(\log(\theta))-b E_{LN}(\theta)\\
& =  a\log b - \log(\Gamma(a)) + (a-1)\mu-b \exp\left(\mu+\sigma^2/2\right).
\end{align*} 

Deriving it with respect to $a$ and $b$ and setting equal to zero, we get that the maximum of $f(a,b)$ is achieve when:
\begin{align*}
a = \sigma^{-2} \mbox{ and } b = \sigma^{-2}\exp\left[-\left(\mu+\frac{\sigma^2}{2}\right)\right],
\end{align*}
for $\sigma^2<1$, in order to guarantee the existence of the mode of the gamma distribution.

\section*{Appendix B: $\cal{NGAL}$ distribution}

%In order to facilitate the notation, let us omitt the index $t$ and call: $\mu={\bf F}_t'{\bf a}_t$, $\eta=a_{\tau}\phi^{-1/2}$ and $\kappa=\phi^{-1}b_{\tau}$ and 
%$c=\phi^{-1}{\bf F}_t'{\bf R}_{t} {\bf F}_t.$ 

This distribution was first presented in \cite{reed2006normal}, who described some of its properties. A random variable $Y$ has a $\cal{NGAL}$ distribution if its characteristic function is equal to
\begin{align}\label{rep2}
\varphi_y(s) & = \left(\frac{\alpha\beta \exp(i\delta s-\gamma s^2/2)}{(\alpha-is)(\beta-is)}\right)^\rho,
\end{align}
where $\alpha$, $\beta$, $\rho$ and $\gamma$ are positive parameters and $\delta$ is real. We write $Y\sim{\cal NGAL}(\delta,\gamma,\alpha,\beta,\rho)$ to indicate that $Y$ follows such a distribution.
Using the cumulants of the distribution it is possible to obtain:%$k_n$ of the distribution defined as:
% \begin{align}
%  \log(\varphi_y(s)) = \sum_{n=1}^\infty{k_n\frac{(it)^n}{n!}}.
% \end{align}
% 
% Using the Maclaurin series expansions for $\log\frac{x}{x-is}$, we get for the NGAL distribution:
% \begin{align*}
% \log(\varphi_y(s)) = is\rho\left(\delta+\frac{1}{\alpha}-\frac{1}{\beta}\right)+\frac{(is)^2}{2!}\rho\left(h+\frac{1}{\alpha^2}+\frac{1}{\beta^2}\right)+\frac{(is)^3}{3!}\rho\left(\frac{2}{\alpha^3}-\frac{2}{\beta^3}\right)+\dots
% \end{align*}
% where $k_r=\rho(r-1)!\left(\delta+\frac{1}{\alpha^r}+(-1)^r\frac{1}{\beta^r}\right).$
%We therefore obtain for example the NGAL mean and variance, given respectively by:
\begin{align*}
E(Y)=\rho\left(\delta+\frac{1}{\alpha}-\frac{1}{\beta}\right) \mbox{ and } Var(Y)=\rho\left(h+\frac{1}{\alpha^2}+\frac{1}{\beta^2}\right).
\end{align*}

Also, the coefficients of kurtosis and skewness are, respectively, given by:
\begin{align*}
\frac{k_4}{k_2^2}=\frac{6(\alpha^4+\beta^4)}{\rho(h\alpha^2\beta^2+\alpha^2+\beta^2)^2} \mbox{ and } \frac{k_3}{k_2^{3/2}}=\frac{2(\beta^3-\alpha^3)}{\rho^{1/2}(h\alpha^2\beta^2+\alpha^2+\beta^2)^2}.
\end{align*}

% Closed-form expressions for the density and cumulative distribution functions of the family of $\cal{NGAL}$ distributions have not been found except when $\rho= 1$. 
% However, they can be obtained numerically using the convolution form or the inversion of the ch.f. In particular, using the convolution to represent the density and the cumulative distribution function we get, respectively:
% \begin{align}
% f(y) = \int_{-\infty}^{\infty}{f_{\epsilon}(y-z)f_{\zeta}(z)dz} \mbox{ and }
% F(y) = \int_{-\infty}^{\infty}{F_{\epsilon}(y-z)f_{\zeta}(z)dz},
% \end{align}
% where $f_{\zeta}(.)$ is the density of the $\cal{GAL}$ distribution, and $f_{\epsilon}(.)$ and $F_{\epsilon}(.)$ are the density and cumulative distribution function of a normal distribution with mean $0$ and variance $c$. Those integrals can be evaluated numerically using for example adaptive quadrature (\cite{piessens2012quadpack}).

\begin{figure}[!hbt]
\begin{center}
\includegraphics[scale=0.55]{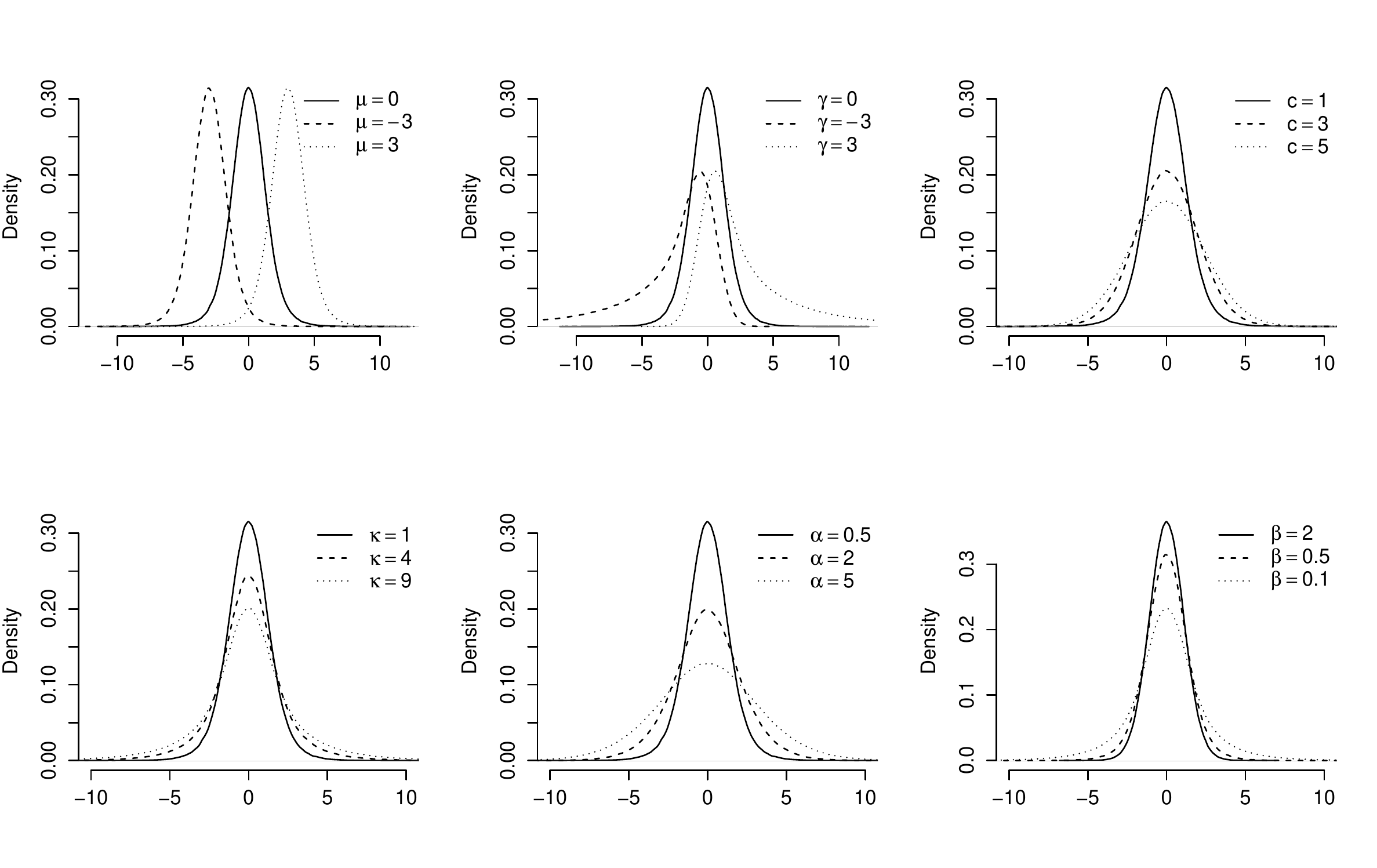}
\vspace{-0.6 cm}\caption{Density function of the $\cal{NGAL}$ distribution with some different values of the parameters.}
\label{densNGAL}
\end{center}
\end{figure}

Figure \ref{densNGAL} compares the density curves of $\cal{NGAL}$ for some values of the parameters.
It is possible to observe that as $\sigma^2$ increases, the density becomes both wider and flatter. The parameters $\alpha$ and $\beta$ affect, 
respectively, the upper and lower tail behavior of the $\cal{NGAL}$ distribution: small values of $\alpha$ and $\beta$ correspond, respectively, to a fat upper and lower tails, while as they increase, the upper and lower tails of the distribution reduce to those of a normal distribution. 
Finally, as $\rho$ increases both mean and variance increase. 

Finally, we need to prove that the distribution described in Theorem \ref{NGAL} is a $\cal{NGAL}$ distribution. In order to facilitate the notation, let us omit the index $t$ and call: $\theta={\bf F}_t'{\bf a}_t$, $\mu=a_{\tau}\phi^{-1/2}\beta_t^{-1}$, $\sigma^2=\phi^{-1}b_{\tau}\beta_t^{-1}$, $c=\phi^{-1}{\bf F}_t'{\bf R}_{t}{\bf F}_t$, 
$\rho=\alpha_t$ and $w=u_t^*$. Thus, we have:
\begin{align*}%\label{mix_repr_forec_appendix}
  y\mid w & \sim N\left(\theta+\mu w, c+\sigma^2 w\right),\\
  w &\sim Ga(\rho,1).
  \end{align*}
  
Conditional on $w$, we obtain the ch.f of $y$ as: 
\begin{align}
\varphi_y(s) &= E\left[E\left(e^{isy}\mid w\right)\right] = e^{is\theta}\int_0^\infty{e^{is\mu w}E\left(e^{is(c+\sigma^2 w)^{1/2}z}\right)}g(w)dw\nonumber
\end{align}

\begin{align}
& = e^{is\theta}\frac{1}{\Gamma(\alpha)}\int_0^\infty{e^{is\mu w}e^{-\frac{1}{2}s^2(c+\sigma^2w)}{w}^{\alpha-1} e^{-w}dw}\nonumber\\
& = e^{is\theta-\frac{1}{2}s^2c}\frac{1}{\Gamma(\alpha)}\int_0^\infty{w^{\alpha-1}e^{-w\left(1+\frac{1}{2}\sigma^2 s^2-i\mu s\right)}dw}\nonumber\\
& = e^{is\theta-\frac{1}{2}s^2c}\left(\frac{1}{1+\frac{1}{2}\sigma^2 s^2-i\mu s}\right)^{\rho}.\label{ch1}
\end{align}

Thus, $\varphi_Y(s) = \varphi_\epsilon(s).\varphi_\zeta(s),$ where $\varphi_\epsilon(s)$ is the N($0,c$) ch.f and $\varphi_\zeta(s)$ is the $\cal{GAL}$($\theta,\mu,\sigma,\rho$) ch.f. 
Thus, it follows that the NGAL distribution is that of the convolution of the normal N($0,c$) and $\cal{GAL}$($\theta,\mu,\sigma,\rho$) distributions, that is $Y=\zeta + \epsilon$, for independent $\zeta$ and $\epsilon$. 

% The ch.f of GAL can be expressed in the following manner:
% \begin{align}\label{ch2}
% \varphi_y(s) & = e^{is\theta}\left(\frac{1}{1+i\frac{\sqrt{2}}{2}\sigma\kappa s}\right)^{\rho}\left(\frac{1}{1-i\frac{\sqrt{2}}{2\kappa}\sigma s}\right)^{\rho},
% \end{align}
% where the additional parameter $\kappa>0$ is related to $\mu$ and $\sigma$ as follows:
% \begin{align}
% \kappa = \frac{\sqrt{2\sigma^2 +\mu^2}-\mu}{\sqrt{2}\sigma} \mbox{, for } \mu=\frac{\sigma}{\sqrt{2}}\left(\frac{1}{\kappa}-\kappa\right).
% \end{align}

Moreover, the ch.f in $(\ref{ch1})$ can be expressed as stated in (\ref{rep2}), for $\alpha=\frac{2\kappa}{\sqrt{2\sigma}}$, $\beta=\frac{2}{\sqrt{2}\sigma\kappa}$, $\delta=\theta/\rho$, $\gamma=c/\rho$ and the additional parameter $\kappa>0$ is related to $\mu$ and $\sigma$ as follows:
\begin{align*}
\kappa = \frac{\sqrt{2\sigma^2 +\mu^2}-\mu}{\sqrt{2}\sigma} \mbox{, while } \mu=\frac{\sigma}{\sqrt{2}}\left(\frac{1}{\kappa}-\kappa\right),
\end{align*}
showing that it is $\cal{NGAL}$ distributed.

\section*{Appendix C: Assessment by MCMC}

\begin{figure}[!hbt]
\begin{center}
\subfigure[$t = 1$]{\includegraphics[scale=0.25]{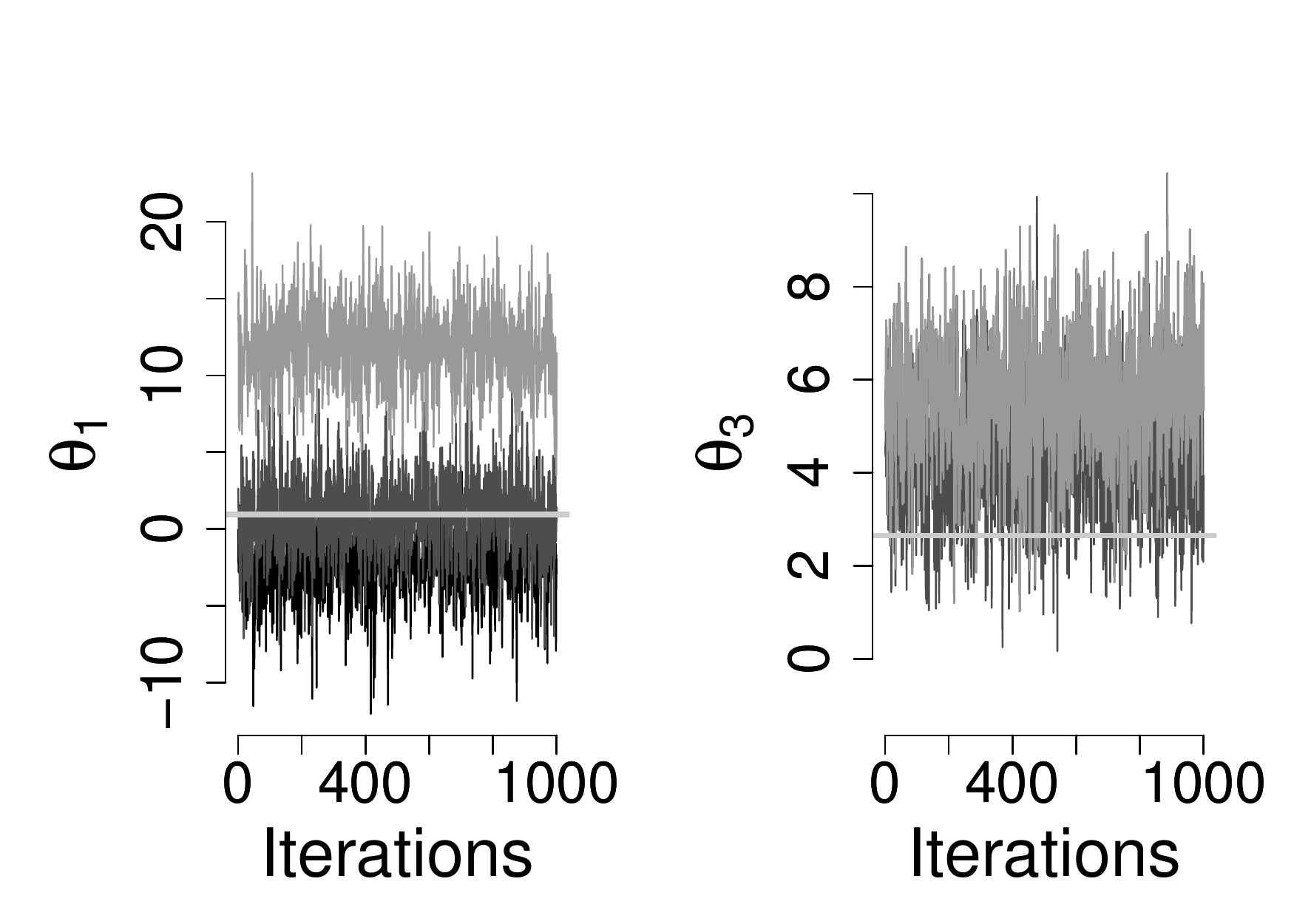}} % salve como 5 por 7 no R
 \subfigure[$t = 10$]{\includegraphics[scale=0.25]{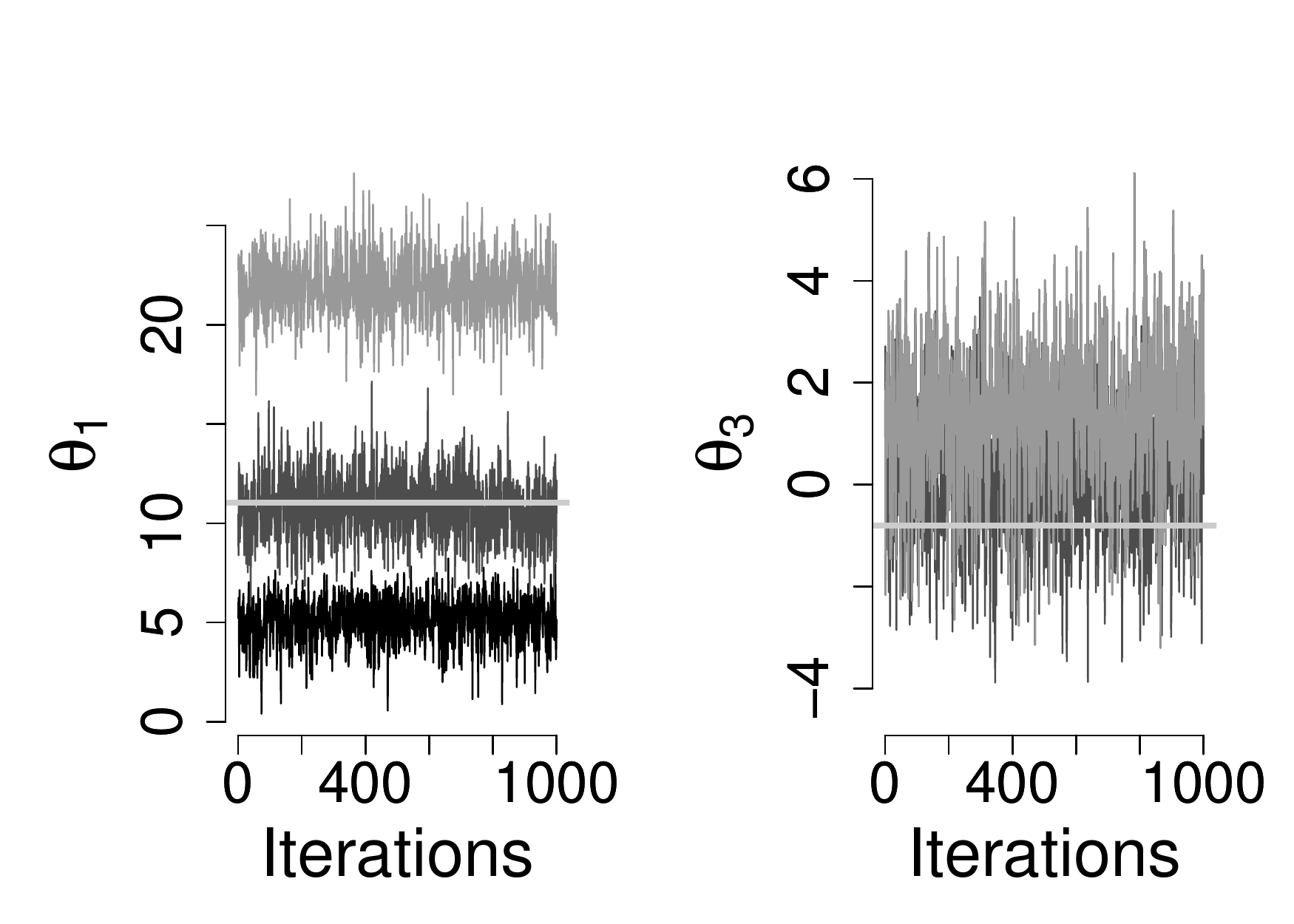}}\\%\vspace{-1 cm}
 \subfigure[$t = 40$]{\includegraphics[scale=0.25]{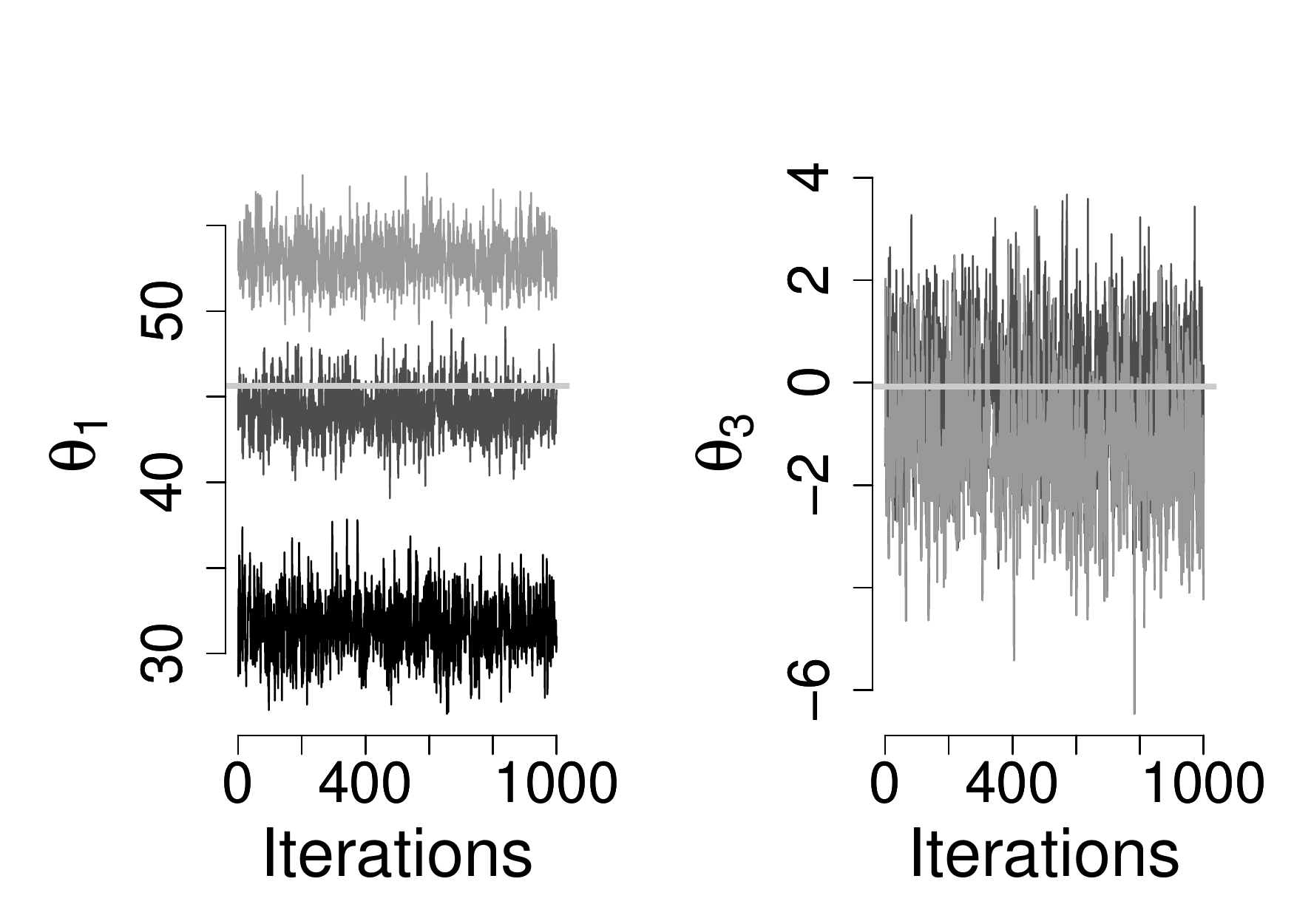}}
 \subfigure[$t = 80$]{\includegraphics[scale=0.25]{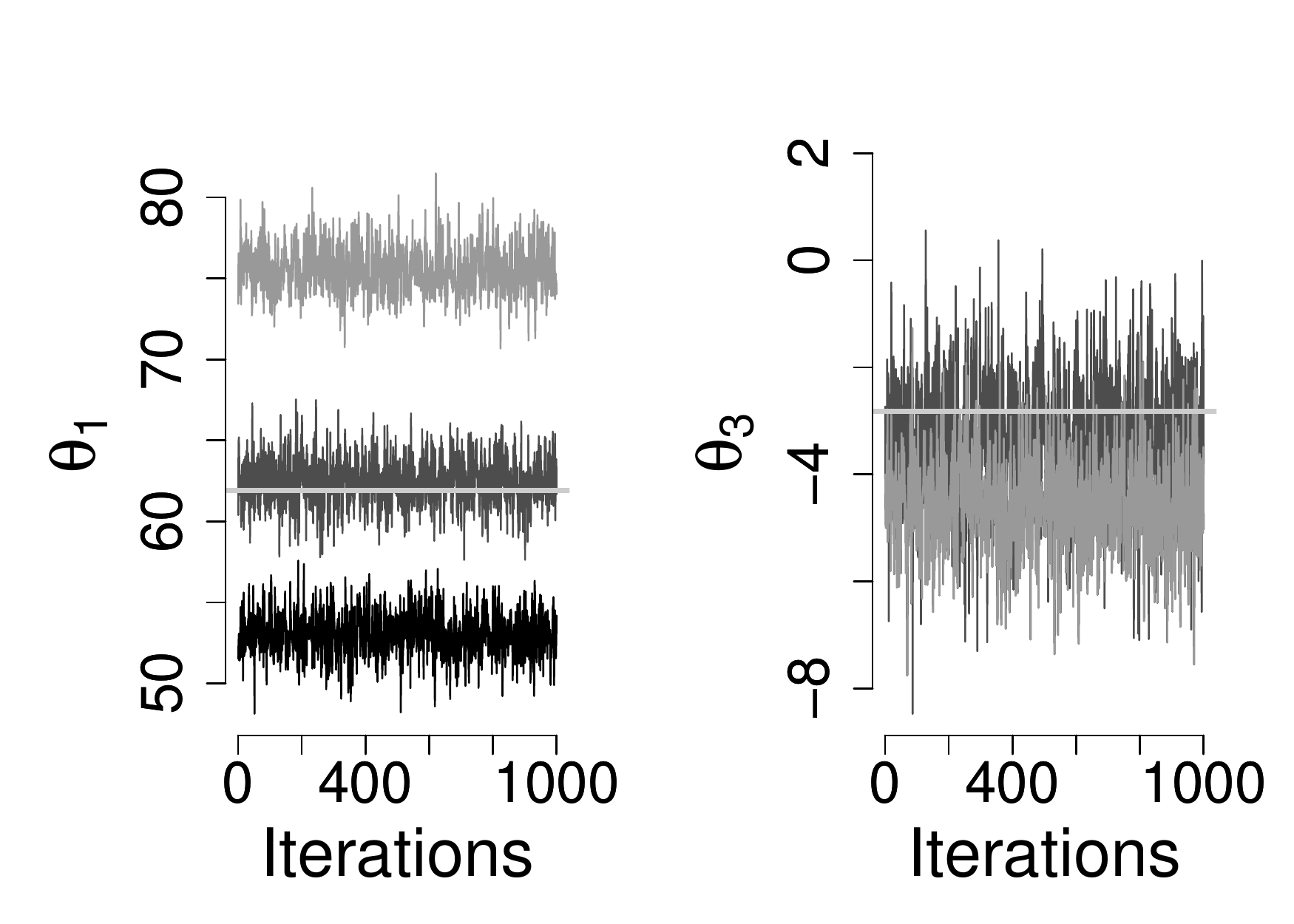}}
\vspace{-0.3 cm}\caption{Trace plot with the posterior densities of $\theta_t$'s for some times and $\sigma$ for each quantile regression fitted.}
\label{chain_saz}
\end{center}
\end{figure}

Figure \ref{chain_saz} shows the trace plot with the posterior distribution of parameters $\bftheta_t$'s for some 
 times for each quantile regression fitted. The chains in black are obtained 
for quantile 0.10, in dark gray for quantile 0.50  and in light gray for  0.90. The line represents the true value of each component of 
$\bftheta_t$ used in the data generating process. 

%\clearpage

\begin{figure}[h!]
\begin{center}
\includegraphics[scale=0.45]{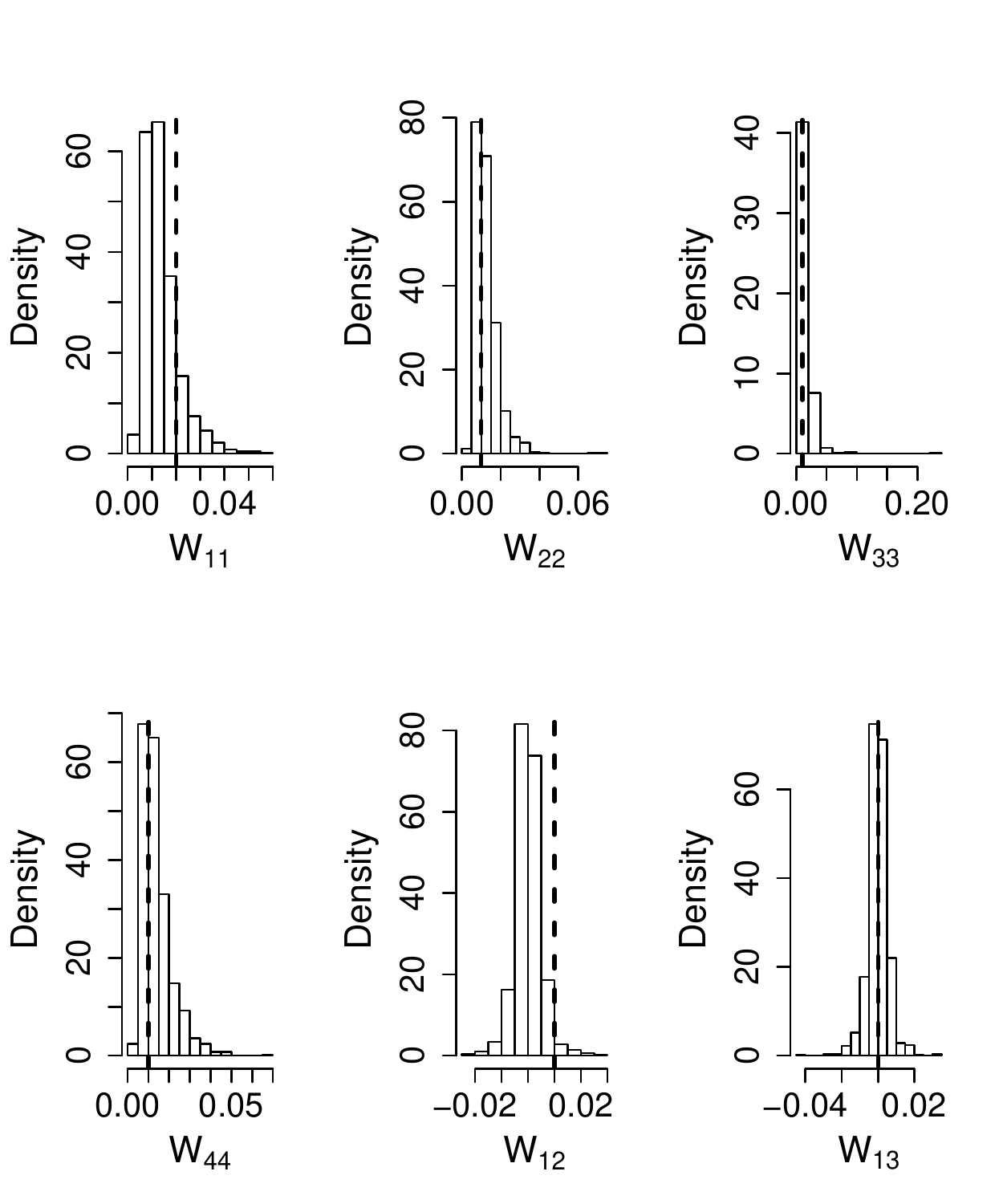} % salvei como 6 por 5 no R
\end{center}
\vspace{-0.8 cm}\caption{Histograms with the posterior densities of the hyperparameters in the
diagonal of the covariance matrix ${\bf W}$.}\label{hist_hyperparam}
\end{figure}

Figure \ref{hist_hyperparam} presents histograms  of  the posterior densities of some elements of 
the covariance  matrix ${\bf W}$, where the true value used in the data generation process is given by the vertical dashed line. 
The hyperparameters appear to be well estimated.

\clearpage

\end{document}